\documentclass[journal]{IEEEtran}

\usepackage{mathtools}
\usepackage{amsmath} 
\usepackage{amssymb}
\usepackage{amsfonts}
\usepackage{verbatim} 
\usepackage{bm,color,soul}
\usepackage{algorithm,algorithmic}
\usepackage{cite}
\usepackage{caption}
\usepackage{subfigure}
\usepackage{stfloats} 
\usepackage[colorlinks, linkcolor=black, anchorcolor=black, citecolor=black]{hyperref} 
\usepackage{graphicx} 
\usepackage{subeqnarray}
\usepackage{cases}
\usepackage{makecell} 
\usepackage{enumerate}
\usepackage{array}
\usepackage{url}
\usepackage{balance}

\newtheorem{rem}{Remark}
\newtheorem{prop}{Proposition}
\newtheorem{lem}{Lemma}

\newcommand{\rnum}[1]{\uppercase\expandafter{\romannumeral #1\relax}}
\ifCLASSINFOpdf
\else
\fi
\makeatletter
\renewcommand{\maketag@@@}[1]{\hbox{\m@th\normalsize\normalfont#1}}
\makeatother

\IEEEoverridecommandlockouts

\begin{document}
	%
	\title{Multi-IRS-Enabled Integrated Sensing and Communications \vspace{-0em} }

	\author{\IEEEauthorblockN{Yuan Fang, Siyao Zhang, Xinmin Li, Xianghao Yu, Jie Xu, and Shuguang Cui
	}\vspace{-1cm}	
		\thanks{Y. Fang is with the Future Network of Intelligence Institute (FNii) and the School of Science and Engineering (SSE), The Chinese University of Hong Kong (Shenzhen), Shenzhen, 518172, China (e-mail: fangyuan@cuhk.edu.cn).} 
		\thanks{S. Zhang is with the School of Computer Science and Technology, Southwest University of Science and Technology, Mianyang 621000, China, and the FNii, The Chinese University of Hong Kong (Shenzhen), Shenzhen, 518172, China (e-mail: zsy@mails.swust.edu.cn).}
		\thanks{X. Li is with the School of Information Engineering, Southwest University of Science and Technology, Mianyang 621000, China, and the FNii, The Chinese University of Hong Kong (Shenzhen), Shenzhen, 518172, China (e-mail: lixm@swust.edu.cn).} 
  \thanks{X. Yu is with the Department of Electrical Engineering, City University of Hong Kong, Hong Kong (e-mail: alex.yu@cityu.edu.hk)}
		\thanks{J. Xu and S. Cui are with the SSE and FNii, The Chinese University of Hong Kong (Shenzhen), Shenzhen, 518172, China (e-mail: xujie@cuhk.edu.cn, shuguangcui@cuhk.edu.cn). X. Li and J. Xu are the corresponding authors.
	}
	}
	
	\maketitle
	\begin{abstract}
		This paper studies a multi-intelligent-reflecting-surface-(IRS)-enabled integrated sensing and communications (ISAC) system, in which multiple IRSs are installed to help the base station (BS) provide ISAC services at  separate line-of-sight (LoS) blocked areas. We focus on the scenario with semi-passive uniform linear array (ULA) IRSs for sensing, in which each IRS is integrated with dedicated sensors for processing echo signals, and each IRS simultaneously serves one sensing target and \textcolor{black}{multiple communication users (CUs)} in its coverage area. In particular, we suppose that the BS sends combined information and dedicated sensing signals for ISAC. Two cases with point and extended targets are considered, in which each IRS aims to estimate the direction-of-arrival (DoA) of the corresponding target and the complete target response matrix, respectively. Under this setup, we first derive the closed-form Cram{\'e}r-Rao bounds (CRBs) for parameters estimation under the two target models. For the point target case, the CRB for DoA estimation is shown to be inversely proportional to the cubic of the number of sensors at each IRS, while for the extended target case, the CRB for target response matrix estimation is proportional to the number of IRS sensors. Next, we consider two different types of CU receivers that can and cannot cancel the interference from dedicated sensing signals prior to information decoding. To achieve fair and optimized sensing performance, we minimize the maximum CRB at all IRSs for the two target cases, via jointly optimizing the transmit beamformers at the BS and the reflective beamformers at the multiple IRSs, subject to the minimum signal-to-interference-plus-noise ratio (SINR) constraints at individual CUs, the maximum transmit power constraint at the BS, and the unit-modulus constraints at the multiple IRSs. To tackle the highly non-convex SINR-constrained max-CRB minimization problems, we propose efficient algorithms based on alternating optimization and semi-definite relaxation, to obtain converged solutions. Finally, numerical results are provided to verify the benefits of our proposed designs over various benchmark schemes based on separate or heuristic beamforming designs.
	\end{abstract}
	

	%
	\IEEEpeerreviewmaketitle

	\section{Introduction}
%
	
	\textcolor{black}{Integrated sensing and communications (ISAC) has been recognized as one of the usage scenarios for future sixth-generation (6G) wireless networks \cite{cui2021integrating,zhang2021overview,liu2022integrated,zhang2022integration},} in which the spectrum resources, wireless infrastructures, and radio signals are reused for the dual role of radar sensing and wireless communications. On the one hand, ISAC allows the cellular base stations (BSs) and mobile devices to jointly design the transmit signals and waveforms to perform both sensing and communications, thus properly managing the co-channel interference between them and accordingly enhancing the resource utilization efficiency. On the other hand, ISAC can also enable the coordination between sensing and communications, such that the sensing information can be exploited to facilitate communications by, e.g., reducing the conventional training overhead, and the communications among different nodes can be utilized to support networked sensing from different views in a large scale \cite{huang2022coordinated}. As such,  ISAC is expected to significantly improve the sensing and communication performances for supporting various new applications such as auto-driving, extended reality, and unmanned aerial vehicles (UAVs) \cite{lyu2022joint}. To make ISAC a reality, extensive research efforts have been devoted to studying  innovative ISAC system designs such as the waveform and beamforming optimization (see \cite{hua2023optimal,liu2022cramer} and the references therein). However, due to the complicated radio propagation environments, the practical ISAC performance may degrade seriously when the transmission links are blocked by obstacles such as trees and buildings. 
	
	Intelligent reflecting surface (IRS) provides a viable solution to enhance both sensing and communication performances by reconfiguring the radio environments via reflecting the incident signals with properly controlled phases and/or amplitudes \cite{huang2019reconfigurable,di2020smart,wu2021intelligent,chu2023joint}. In particular, IRSs can be used in wireless communications networks to help enhance the signal coverage, increase the desired signal strength, mitigate the undesired interference, reshape the communication channel rank, and accordingly boost the communication performance \cite{xie2020max,pan2022overview,li2023beyond}. \textcolor{black}{Furthermore, IRSs can also be employed in wireless sensing systems to create virtual line-of-sight (LoS) links to see targets located in non-line-of-sight (NLoS) regions, provide multi-view sensing for facilitating the target detection and estimation\textcolor{black}{\cite{hu2022irs}}, and enable networked target localization \cite{zhang2020perceptive,buzzi2022foundations,jiang2023two}.}
	 
	 While there have been extensive works investigating IRS-enabled wireless communications \cite{liu2023integrated}, there have been only a handful of prior works studying IRS-enabled sensing \cite{shao2022target,hua2023intelligent,song2022intelligent} and IRS-enabled ISAC \cite{wang2021jointTVT,song2022joint,liu2022joint,hua2022joint}. \textcolor{black}{First, the IRS-enabled sensing can be generally implemented in two different modes, namely the semi-passive IRS sensing \cite{shao2022target,hua2023intelligent} and passive IRS sensing \cite{song2022intelligent}, where the IRS is installed with and without the dedicated sensors for signal reception, respectively. The former mode processes the sensing signals at the IRS directly, while the latter requires remote signal processing at the BS. Compared to passive IRS sensing, semi-passive IRS sensing reduces the propagation path loss since the semi-passive IRS does not need to further reflect echo signals from the IRS to BS.}  In particular, the authors in \cite{song2022intelligent} studied the passive IRS sensing for estimating the target angle with respect to the IRS based on the echo signals over the BS-IRS-target-IRS-BS link. In \cite{song2022intelligent}, the authors first analyzed the Cram{\'e}r-Rao bounds (CRBs) for the BS to estimate the angle of the point target or the target response matrix of the extended target, and then proposed to jointly optimize the active transmit beamforming and the passive reflective beamforming to minimize the CRBs. \textcolor{black}{In contrast, \cite{shao2022target} and \cite{hua2023intelligent} studied the semi-passive IRS sensing for localizing the targets by using the echo signals over the BS-IRS-target-IRS link, in which the IRS passive reflection matrix was optimized to maximize its received echo signal power and a low-complexity location estimation algorithm was further proposed.} Next, the works \cite{wang2021jointTVT,song2022joint,liu2022joint,hua2022joint} considered the IRS-enabled ISAC, among which \cite{wang2021jointTVT} studied the scenario with an IRS assisting communications only and \cite{song2022joint,liu2022joint,hua2022joint} considered passive IRS sensing. In particular, by considering the passive IRS sensing, the work \cite{song2022joint} maximized the minimum reflected beampattern gains towards multiple sensing angles with respect to the IRS by jointly optimizing the transmit and reflective beamforming. In \cite{liu2022joint}, the authors studied a scenario where the target sensing is corrupted by multiple clutters with signal-dependent interference, in which the radar signal-to-interference-plus-noise ratio (SINR) was maximized. The work \cite{hua2022joint} proposed to jointly design the communication and sensing waveform to minimize the total transmit power at BS while ensuring the constraints on both communication and radar SINR as well as the cross-correlation pattern design requirements. However, the above prior works on IRS-enabled sensing and IRS-enabled ISAC mainly focused on the case with one single IRS, which has a limited coverage area due to the round-trip path loss. In practice, each BS may have multiple sensing coverage holes due to the distributed obstructions. Therefore, it is important to exploit multiple IRSs to cooperate in helping BS cover different LoS blocked areas. 
	
	 In the literature, the multi-IRS-enabled wireless communications have been extensively studied, in which multiple IRSs are deployed for enhancing the coverage areas via either independent single-hop reflections \cite{li2019joint,yang2021energy,nguyen2022leveraging} or cooperative multi-hop reflections \cite{huang2021multi,liang2022tvt,mei2020cooperative,mei2021multi}. For instance, the authors in  \cite{li2019joint}  studied a downlink multi-IRS multi-user communication system with each IRS serving one user, in which the joint beamforming was optimized to maximize the minimum SINR among all users. The work \cite{yang2021energy} studied a downlink wireless communication system with multiple distributed IRSs, in which the on-off control at the IRSs were optimized jointly with the resource allocation to maximize the system energy efficiency while ensuring the minimum achievable rate requirements at the users. The authors in \cite{nguyen2022leveraging} further investigated the impact of secondary reflections in an uplink multi-IRS-aided multi-user communication system, in which the passive beamforming at these IRSs were optimized to maximize the weighted-sum rate or the minimum rate at multiple users. Furthermore, the authors in \cite{huang2021multi} studied the IRS-enabled multi-refection between a BS and multiple users, in which the active beamforming at the BS and the passive beamforming at multiple IRSs were jointly designed based on deep reinforcement learning. In addition, \cite{liang2022tvt} investigated a multi-route multi-hop cascaded IRS communication system, with the objective of maximizing the achievable sum rate at the multiple users via the joint design of active and cascaded passive beamforming. In \cite{mei2020cooperative} and \cite{mei2021multi}, the authors studied the optimal multi-IRS-reflection path selection for multi-IRS beam routing problem. On the other hand, multiple IRSs can be employed to enhance the signal coverage of BS via single reflection of each IRS. Despite the research progress, the above prior works only studied multi-IRS-enabled wireless communications. To our best knowledge, how to efficiently implement multi-IRS-enabled sensing and multi-IRS-enabled ISAC has not been well investigated in the literature yet. This thus motivates our investigation in this work.
	 
This paper studies the multi-IRS-enabled ISAC system, in which multiple semi-passive IRSs are deployed at different locations to assist the BS to provide seamless ISAC services via their single reflections.\footnote{How to use multi-reflection or multi-hop links of multiple IRSs for sensing and ISAC is also an interesting direction, which, however, is beyond the scope of this paper and is left for future work.} \textcolor{black}{Different from previous works \cite{hua2023optimal} and \cite{song2022intelligent}, which studied a MIMO ISAC system without IRS and a single-passive-IRS-enabled sensing system, respectively, the multi semi-passive-IRS-enabled ISAC studied in this paper is able to enhance the coverage areas of both sensing and communications with reduced hardware and operational costs.} However, as compared to the single-IRS-enabled counterpart, the multi-IRS-enabled ISAC introduces new technical challenges due to the involvement of multiple IRSs. \textcolor{black}{In particular, as the transmit signals from the BS need to be reflected by different IRSs to serve their respective areas, the derived estimation CRB with a single passive-IRS \cite{song2022intelligent} cannot be directly used for the multi-semi-passive-IRS scenario. Furthermore, as there are multiple IRSs associated with one BS, the joint design of transmit beamforming at the BS and reflective beamforming at the IRS proposed for single-passive-IRS-enabled sensing or ISAC in the existing works is inapplicable. How to coordinate the reflective beamforming at all IRSs together with the transmit optimization at the BS is also a paramount yet difficult task.}

More specifically, this paper investigates the multi-IRS-enabled ISAC system consisting of one BS and multiple semi-passive IRSs each equipped with a uniform linear array (ULA), \textcolor{black}{in which each IRS employs reflective beamforming to serve multiple communication users (CUs) and sense one target at the same time.} To facilitate ISAC, the BS sends dedicated sensing signals combined with information signals. As the dedicated sensing signals can be known {\it a priori},  we consider two different types of CU receivers that have and do not have the capability of cancelling the interference caused by the sensing signals, respectively.  Besides, we also consider two cases with point and extended targets, for which each IRS aims to estimate the direction-of-arrival (DoA) of its corresponding target, and the complete response matrix between each IRS and its corresponding target, respectively. The main results of this paper are summarized as follows.
\begin{itemize}
\item First, different from most prior works on IRS-enabled ISAC that used the transmit beampattern and radar SINR as the sensing performance measures, we consider the CRB for target estimation as the metric to characterize the fundamental sensing performance, which serves as the variance lower bound of any practical biased estimators. \textcolor{black}{In particular, for the point and extended targets, we derive the closed-form CRBs for estimating the target DoA and the complete target response matrix, respectively. The derived CRB for DoA estimation is shown to be inversely proportional to the cubic of the number of sensors at each IRS for the point target case, and the CRB for target response matrix estimation is proportional to the number of sensors at each IRS for the extended target case.}
\item \textcolor{black}{Next, under the two target models and by considering two different CU types, we minimize the maximum CRB for targets estimation at all IRSs to achieve a fair and efficient sensing performance. In particular, the transmit beamformers at the BS and the reflective beamformers at multiple semi-passive IRSs are jointly optimized, subject to the minimum SINR constraints at individual CUs, the maximum transmit power constraint at the BS for transmit beamforming, and the unit-modulus constraints at the IRSs for reflective beamforming.} The four SINR-constrained max-CRB minimization problems are highly non-convex due to the coupling of transmit beamforming at BS and passive beamforming at IRSs as well as the unit-modulus constraints. To tackle these problems, we propose efficient algorithms to obtain converged solutions based on alternating optimization and semi-definite relaxation (SDR).
\item \textcolor{black}{Finally, numerical results show that our proposed designs perform close to the sensing performance upper bound (or CRB lower bound) with sensing task only, especially when the SINR constraints become loose. Moreover, the proposed designs outperform various benchmark schemes based on transmit beamforming only or zero-forcing (ZF) beamforming.} It is also shown that the sensing signal interference cancellation is essential in further enhancing the ISAC performance.
\end{itemize}

	The rest of the paper is organized as follows. Section II presents the multi-semi-passive-IRS-enabled  ISAC system model. Section III derives the closed-form estimation CRBs for both point and extended target cases. Section IV and Section V present the joint transmit and reflective beamforming solutions to the SINR-constrained max-CRB minimization problems for the cases with point and extended targets, respectively. Section VI presents numerical results to validate the performance of our proposed joint beamforming designs as compared to other benchmarks. Finally, Section VII concludes this paper.
	
	{\it Notations:} The circularly symmetric complex Gaussian (CSCG) distribution with mean $\boldsymbol{\mu}$ and covariance $\boldsymbol{A}$ is denoted as $\mathcal{CN}(\boldsymbol{\mu},\boldsymbol{A})$. The notations $(\cdot)^{T}$, $(\cdot)^{*}$, $(\cdot)^{H}$, and $\mathrm{tr}(\cdot)$ denote the transpose, conjugate, conjugate-transpose, and trace operators, respectively. $\boldsymbol{I}_{L}$ stands for the identity matrix of size $L \times L$. $\mathrm{Re}(\cdot)$ and $\mathrm{Im}(\cdot)$ denote the real and imaginary parts of the argument, respectively. $|\cdot|$ and $\mathrm{arg}\left\{\cdot\right\}$ denote the
	absolute value and angle of a complex element, respectively. $\mathrm{vec}(\cdot)$ denotes the vectorization operator, $\mathbb{ E}(\cdot)$ denotes the expectation operation, and $\mathrm{diag}(\boldsymbol{x})$ denotes a diagonal matrix with the diagonal entries specified by vector $\boldsymbol{x}$. ${\rm rank}\left(\boldsymbol{X}\right)$ denotes the rank value of matrix $\boldsymbol{X}$ and $[\cdot]_{l,p}$ denotes the $(l,p)$-th element of a matrix. $j$ denotes the imaginary unit. $\mathcal{A}\backslash a$ denotes the set after removing the element $a$ in $\mathcal{A}$.

	\section{System Model}
	\begin{figure}[tbp]
		\vspace{-5pt}
		\setlength{\abovecaptionskip}{-0pt}
		\setlength{\belowcaptionskip}{-15pt}
		\centering
		\includegraphics[width=0.45\textwidth]{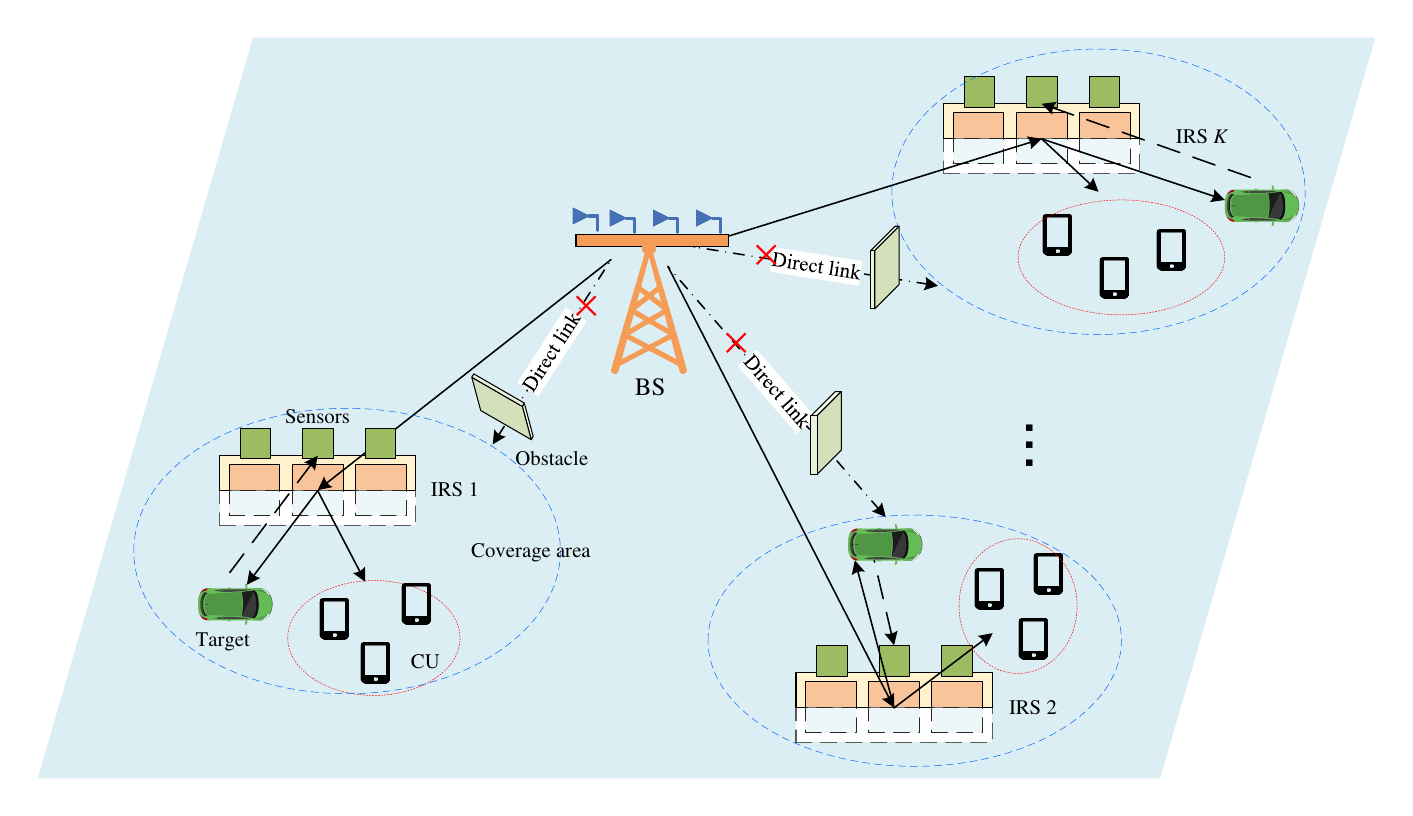}
		\caption{\textcolor{black}{The multi-semi-passive-IRS-enabled ISAC system.}}\label{fig:SystemModel}
	\end{figure}	
	We consider a multi-IRS-enabled ISAC system as shown in Fig. \ref{fig:SystemModel}, which consists of an ISAC BS with $M$ antennas and $L$ semi-passive IRSs. Each IRS is equipped with a ULA of $N$ reflecting elements and $N_s$ receive antenna elements for sensing. Let $\mathcal{L} = \{1,\ldots,L\}$ denote the set of IRSs and $\mathcal{N} = \{1,\ldots,N\}$ denote the set of elements at each IRS. In practice, the coverage of each IRS is limited, and thus we assume that each IRS is deployed to provide ISAC coverage at one separate area by ignoring the interference across different areas \cite{sankar2022beamforming}. Furthermore, it is assumed that there are one sensing target and $K$ signal-antenna CU at the coverage area of each IRS, and the direct links between the BS and the targets/CUs are blocked as assumed in prior work \cite{shao2022target}. Let $\mathcal{K}_{l} = \{1,\ldots, K\}$ denote the set of users at the coverage of IRS $l$.
	
	We consider the quasi-static channel model, in which the wireless channels remain unchanged over the transmission block of interest. Let $T$ denote the block duration or the radar dwell time. Let $\boldsymbol{G}_{l} \in \mathbb{C}^{N \times M}$ denote the channel matrix between the BS and IRS $l$,  and $\boldsymbol{h}_{l,k} \in \mathbb{C}^{N\times 1}$ denote the channel vector between IRS $l$ and the CU $k$ at its coverage area, which are assumed to be known by the system via proper channel estimation (see, e.g., \cite{zheng2022survey}). 
	
	We consider the transmit beamforming at the BS and the reflective beamforming at the IRSs. In particular, the BS sends combined information and sensing signals to facilitate ISAC \cite{hua2023optimal,liu2022cramer}. Let $s_{l,k}[t]$ denote the information signal for CU $k$ in the coverage area of IRS $l$ at time symbol $t$, and $\boldsymbol{w}_{l,k}\in \mathbb{C}^{M\times 1}$ denote the corresponding transmit beamforming vector. Then, the information signal vector for all CUs is denoted as $\boldsymbol{s}[t] = \left[s_{1,1}[t],\ldots,s_{1,K}[t],\ldots,s_{L,1}[t],\ldots,s_{L,K}[t]\right]^T$ with $\boldsymbol{s}[t] \sim \mathcal{CN}(\boldsymbol{0},\boldsymbol{I}_{LK})$. Let $\boldsymbol{s}_{0}[t] \in \mathbb{C}^{M\times 1}$ denote the dedicated sensing signal, which is randomly generated independent of $\boldsymbol{s}[t]$, with zero mean and covariance matrix $\boldsymbol{R}_{{0}}=\mathbb{E}(\boldsymbol{s}_{0}[t]\boldsymbol{s}_{0}^{H}[t])\approx \frac{1}{T} \sum_{t=1}^{T}\boldsymbol{s}_{0}[t]\boldsymbol{s}_{0}^{H}[t]\succeq \boldsymbol{0}$.\footnote{The statistical and sample covariance matrices are assumed to be approximately the same by considering $T$ to be sufficiently large.}
By combining the information and sensing signals,  the transmit signal at the BS is $\boldsymbol{x}[t] = \sum\nolimits_{l \in \mathcal{L}}\sum\nolimits_{k \in \mathcal{K}_{l}}\boldsymbol{w}_{l,k}{s}_{l,k}[t] + \boldsymbol{s}_{0}[t]$, for which the covariance matrix is given by $\boldsymbol{R}_{x} = \sum\nolimits_{l \in \mathcal{L}}\sum\nolimits_{k \in \mathcal{K}_{l}}\boldsymbol{W}_{l,k}+ \boldsymbol{R}_{0}$, where $\boldsymbol{W}_{l,k} = \boldsymbol{w}_{l,k}\boldsymbol{w}_{l,k}^{H}$ with ${\rm rank}\left(\boldsymbol{W}_{l,k}\right) = 1$ and $\boldsymbol{W}_{l,k}  \succeq 0$.
	By letting  $P_{\text{sum}}$ denote the maximum transmit power at the BS, then we have the maximum transmit power constraint at the BS as $\mathbb{E}\left(\|\boldsymbol{x}[t]\|^2\right)  =\mathrm{tr}\left(\boldsymbol{R}_{x} \right) = \mathrm{tr}\left(\sum\nolimits_{l \in \mathcal{L}}\sum\nolimits_{k \in \mathcal{K}_{l}}\boldsymbol{W}_{l,k}+ \boldsymbol{R}_{0}\right) \leq P_{\text{sum}}$. Furthermore, we consider that each IRS $l$ can adaptively adjust the reflection coefficients of its elements based on the channel conditions. Let $\boldsymbol{\phi}_{l} = [\phi_{l,1},\ldots,\phi_{l,n},\ldots, \phi_{l,N}]^{T}$ denote the complex reflection coefficients imposed by IRS $l$, where $|\phi_{l,n}|=1$ and $\mathrm{arg}\left\{\phi_{l,n}\right\} \in (0,2\pi]$, $\forall n \in \{1,\ldots,N\}$. 
	
First, we consider the wireless communication from the BS to the CUs assisted by the IRSs. By omitting the direct link from the BS to each CU $k$ at the coverage area of IRS $l$, the received signal by the CU $k$ at the coverage of IRS $l$  through the BS-IRS-CU link at time symbol $t$ is
	\begin{align}
		{y}_{l,k}[t] &= \boldsymbol{h}_{l,k}^{H}\boldsymbol{\Phi}_{l}\boldsymbol{G}_{l}\boldsymbol{x}[t]+{z}_{l,k}[t]\nonumber\\
		&=\underbrace{\boldsymbol{h}_{l,k}^{H}\boldsymbol{\Phi}_{l}\boldsymbol{G}_{l}\boldsymbol{w}_{l,k}s_{l,k}[t]}_{\text{User's desired signal}}\nonumber\\
		&+\underbrace{\sum\limits_{l' \in \mathcal{L} }\sum\limits_{k' \in \mathcal{K}_{l'} \backslash k} \!\!\!\!\boldsymbol{h}_{l,k}^{H}\boldsymbol{\Phi}_{l}\boldsymbol{G}_{l}\boldsymbol{w}_{l',k'}s_{l',k'}[t]}_{\text{Inter-user interference}}\nonumber\\
		& +  \underbrace{\boldsymbol{h}_{l,k}^{H}\boldsymbol{\Phi}_{l}\boldsymbol{G}_{l} \boldsymbol{s}_{0}[t]}_{\substack{\text{Dedicated sensing} \\ \text{signal interference}}} +{z}_{l,k}[t],
	\end{align}
	where $\boldsymbol{\Phi}_{l} = \text{diag}(\boldsymbol{\phi}_{l})$, and ${z}_{l,k}[t]\sim \mathcal{CN}(0,\sigma^2) $ denotes the additive white Gaussian noise (AWGN) at each CU that may include the background interference among different IRSs. As the sensing signal $\boldsymbol{s}_{0}[t]$ can be generated offline, it can be known prior to transmission. As such, we consider two different types of CU receivers depending on whether they have the capability of cancelling the interference by sensing signals  $\boldsymbol{s}_{0}[t]$.
	\begin{itemize}
		\item \textbf{Type-I CU receiver:} Such CU receivers do not have the capability of canceling the interference by $\boldsymbol{s}_{0}[t]$ and thus treat such interference as noise. In this case, the received SINR of CU $k$ given by \eqref{sinr_I} the top of this page.
		\begin{figure*}
		\begin{align}
		&\!\!\!\!\!\!\!\!\!\!\!\!\gamma_{l,k}^{(\text{I})} (\{\boldsymbol{W}_{l,k}\},\{\boldsymbol{\Phi}_{l}\},\boldsymbol{R}_{0}) \!=\!\frac{{{ \boldsymbol{h}_{l,k}^{H}\boldsymbol{\Phi}_{l}\boldsymbol{G}_{l}\boldsymbol{W}_{l,k}\boldsymbol{G}_{l}^{H}\boldsymbol{\Phi}_{l}^{H}\boldsymbol{h}_{l,k} }}}{{\sum\limits_{l' \in \mathcal{L} }\sum\limits_{k' \in \mathcal{K}_{l'} \backslash k}\!\!\!\!\!\! {{\boldsymbol{h}_{l,k}^{H}\boldsymbol{\Phi}_{l}\boldsymbol{G}_{l}\boldsymbol{W}_{l',k'}\boldsymbol{G}_{l}^{H}\boldsymbol{\Phi}_{l}^{H} \boldsymbol{h}_{l,k} }} \!\!+\!\!\boldsymbol{h}_{l,k}^{H}\boldsymbol{\Phi}_{l}\boldsymbol{G}_{l} \boldsymbol{R}_{0}\boldsymbol{G}_{l}^{H} \boldsymbol{\Phi}_{l}^{H}\boldsymbol{h}_{l,k}\!\!+\!\! \sigma^2}}.\label{sinr_I}\\
		&\gamma_{l,k}^{(\text{II})} (\{\boldsymbol{W}_{l,k}\},\{\boldsymbol{\Phi}_{l}\}) = \frac{{{\boldsymbol{h}_{l,k}^{H}\boldsymbol{\Phi}_{l}\boldsymbol{G}_{l}\boldsymbol{W}_{l,k}\boldsymbol{G}_{l}^{H}\boldsymbol{\Phi}_{l}^{H}\boldsymbol{h}_{l,k}}}}{{\sum\limits_{l' \in \mathcal{L} }\sum\limits_{k' \in \mathcal{K}_{l'} \backslash k} {{\boldsymbol{h}_{l,k}^{H}\boldsymbol{\Phi}_{l}\boldsymbol{G}_{l}\boldsymbol{W}_{l',k'}\boldsymbol{G}_{l}^{H}\boldsymbol{\Phi}_{l}^{H} \boldsymbol{h}_{l,k}}} + \sigma^2}}.\label{sinr_II}
		\end{align}
	\hrulefill	
 \vspace{-15pt}
	\end{figure*}
		\item \textbf{Type-II CU receiver:}
		Such CU receivers are dedicated designed for ISAC, which can cancel the interference from $\boldsymbol{s}_{0}[t]$ before decoding information signal $s_{l,k}[t]$. In this case, the received SINR of CU $k$ is given by \eqref{sinr_II} the top of this page.
	\end{itemize}
	
Next, we consider the target estimation at the IRSs. In particular, we consider two cases with point and extended targets, which are introduced in detail as follows. 
\subsubsection{Point Target Case}
Each point target is assumed to have a small spatial range and can be viewed as a far apart unstructured point for the IRS. In this case, the round-trip target response matrix from IRS $l$ to its corresponding target to IRS $l$ is given by
\begin{align}
	\hat{\boldsymbol{E}}_{l} = \beta_{l}\tilde{\boldsymbol{a}}_{l}(\theta_l)\boldsymbol{a}_{l}^{T}(\theta_l),\label{Point_target_RM}
\end{align} 
where $\beta_l$ denotes the complex coefficient accounting for the radar cross-section of the target and the round-trip path-loss, $\boldsymbol{a}_{l}(\theta_l)$ denotes the array steering vector of reflecting elements at IRS $l$ with DoA $\theta_l$ and $\tilde{\boldsymbol{a}}_{l}(\theta_l)$ denotes the steering vector at the sensors of IRS $l$. Here, we have 
\begin{align}
	\boldsymbol{a}_{l}(\theta_l) &= [1,e^{j \frac{2\pi}{\lambda}d \sin(\theta_l)},\ldots,e^{j\frac{2\pi}{\lambda} (N-1)d\sin(\theta_l)}]^T,\\
	\tilde{\boldsymbol{a}}_{l}(\theta_l) &= [1,e^{j \frac{2\pi}{\lambda}d_s \sin(\theta_l)},\ldots,e^{j\frac{2\pi}{\lambda} (N_s-1)d_s\sin(\theta_l)}]^T,
\end{align}
where $\lambda$ denotes the carrier wavelength, and $d$ and $d_s$ denote the spacing of consecutive reflection elements and sensor elements at IRS $l$, respectively. In this case, the complex coefficient $\beta_l$ and the target DoA $\theta_l$ are unknown parameters to be estimated at each IRS $l$.  
\subsubsection{Extended Target Case}
Extended target refers to an object or an entity that occupies an area rather than being represented as a single point \cite{chen2009mimo}. Unlike a point target that is typically modeled as a single point in space, an extended target can be an object with non-negligible size, such as a vehicle or a group of targets \cite{koch2008bayesian}. The echo signal from an extended target consists of multiple scatterers within the target volume and varies depending on its size, shape, orientation, and material properties. As a result, the point target model in \eqref{Point_target_RM} only reflects a single point scatterer behavior, but is not suitable to characterize extended targets. For tractable theoretical analysis, $\hat{\boldsymbol{E}}_{l}\in \mathbb{C}^{N_s \times N}$ is used to denote the complete target response matrix from IRS $l$ to extended target to IRS $l$, which reflects the comprehensive effect of all scatterers within the target volume.\footnote{\textcolor{black}{Note that for the extended target case, the number of scatterers as well as their distribution are generally unknown in advance. It is thus infeasible to directly estimate the specific scatterer parameters such as their angle and radar cross-section information, as in  the case with point targets.}} In this case, the complete target response matrix $\hat{\boldsymbol{E}}_{l}$ corresponds to the parameters to be estimated at each IRS $l$. \textcolor{black}{After obtaining $\hat{\boldsymbol{E}}_{l}$, the specific parameters of the scatterers such as angel information can be extracted by using some advanced techniques, including multiple signal classification (MUSIC), estimating signal parameters via rotational invariance techniques (ESPRIT), and other learning-based methods \cite{li1996adaptive,roy1989esprit,chen2018off}.}

Based on the two target models, we consider the target sensing at each semi-passive IRS $l$ by using its dedicated sensors for receiving the echo signals reflected by the targets \cite{shao2022target}. In particular, the received echo signal at the sensors of each IRS $l$ from target $l$ at time symbol $t$ is 
	\begin{align}
		\bar{\boldsymbol{y}}_{l}[t] =\hat{\boldsymbol{E}}_{l}\boldsymbol{\Phi}_{l}\boldsymbol{G}_{l}\boldsymbol{x}[t] + \bar{\boldsymbol{z}}_{l}[t],
	\end{align}
	where $\bar{\boldsymbol{z}}_{l}[t] \sim \mathcal{CN}(\boldsymbol{0},\sigma_{{s}}^2\boldsymbol{I}_{N_s})$ denotes the AWGN at the sensor receiver of IRS $l$. By defining $\boldsymbol{X} = [\boldsymbol{x}[1],\ldots,\boldsymbol{x}[T]]$, $\bar{\boldsymbol{Y}}_{l} = [\bar{\boldsymbol{y}}_{l}[1],\ldots,\bar{\boldsymbol{y}}_{l}[T]]$, and $\bar{\boldsymbol{Z}} = [\bar{\boldsymbol{z}}_{l}[1],\ldots,\bar{\boldsymbol{z}}_{l}[T]]$, we have 
	\begin{align}\label{EchoSigAtIRS}
		\bar{\boldsymbol{Y}}_{l} = \hat{\boldsymbol{E}}_{l}\boldsymbol{\Phi}_{l}\boldsymbol{G}_{l}\boldsymbol{X}+ \bar{\boldsymbol{Z}}_{l}.
	\end{align} 
	We suppose that each IRS $l$ is aware of the channel matrix $\boldsymbol{G}_{l}$ and the BS's transmitted signal $\boldsymbol{X}$ via proper channel estimation and signalling from the BS. Accordingly, based on the received echo signal $\bar{\boldsymbol{Y}}_{l}$ in \eqref{EchoSigAtIRS},  each IRS $l$ needs to estimate the DoA $\theta_{l}$ and the complex coefficient $\beta_{l}$ as unknown parameters for the point target case, and needs to estimate the complete target response matrix $\hat{\boldsymbol{E}}_{l}$ for the extended target case. 
	%
	\section{Estimation CRB Derivation}
	In this section, we derive the estimation CRB as the sensing performance metric for each IRS $l$ to estimate the unknown parameters based on \eqref{EchoSigAtIRS}, which provides the variance lower bound for any unbiased estimators.  
\subsection{Point Target Case}
	For the point target case, the target response matrix $\hat{\boldsymbol{E}}_{l}$ is particularly specified by \eqref{Point_target_RM}. Thus, the echo signal received at IRS $l$ in \eqref{EchoSigAtIRS} is re-expressed as
	\begin{align}\label{Response_matrix_irs}
		\bar{\boldsymbol{Y}}_{l} = \beta_{l}\boldsymbol{E}_{l}\boldsymbol{\Phi}_{l}\boldsymbol{G}_{l}\boldsymbol{X}+ \bar{\boldsymbol{Z}}_{l},
	\end{align} 
	where $\boldsymbol{E}_{l} = \tilde{\boldsymbol{a}}_{l}(\theta_l)\boldsymbol{a}_{l}^{T}(\theta_l)$. 
	Let $\boldsymbol{\xi}_{l}  = [\theta_{l},\boldsymbol{\beta}_{l}]^{T}$ denote the three real parameters to be estimated, where $\boldsymbol{\beta}_{l} = [\mathrm{Re}\{\beta_{l}\},\mathrm{Im}\{\beta_{l}\}]$. By vectorizing $\bar{\boldsymbol{Y}}_{l}$ in \eqref{Response_matrix_irs}, we have 
	\begin{align}
		\tilde{\boldsymbol{y}}_{l} = \mathrm{vec}(\bar{\boldsymbol{Y}}_{l}) = \tilde{\boldsymbol{\eta}}_{l} + \tilde{\boldsymbol{z}}_{l},\label{Vec_Y}
	\end{align}
	where $\tilde{\boldsymbol{\eta}}_{l} = \mathrm{vec}\left(\beta_{l}\boldsymbol{E}_{l}\boldsymbol{\Phi}_{l}\boldsymbol{G}_{l}\boldsymbol{X}\right)$ and $\tilde{\boldsymbol{z}}_{l} = \mathrm{vec} (\boldsymbol{Z}_{l}) \sim \mathcal{CN}(\boldsymbol{0},\sigma_{{s}}^2\boldsymbol{I}_{N_s T})$. 
	
	First, we derive the Fisher information matrix (FIM) $\boldsymbol{J}_{l} \in \mathbb{C}^{3 \times 3}$ for estimating the vector $\boldsymbol{\mu}$ from \eqref{Vec_Y}, which is given by \cite{kay1993fundamentals}
	\begin{align}\label{FIM_def_2}
		[\boldsymbol{J}_{l}]_{q_1,q_2} = \frac{2}{\sigma_{{s}}^{2}}\mathrm{Re}\left\{\frac{\partial \tilde{\boldsymbol{\eta}}_{l}^{H}}{\partial [\boldsymbol{\xi}_{l}]_{q_1}} \frac{\partial \tilde{\boldsymbol{\eta}_{l}}}{\partial [\boldsymbol{\xi}_{l}]_{q_2}} \right\}, q_1,q_2 \in \{1,2,3\}.
	\end{align}
Recall that the statistic covariance matrix $\boldsymbol{R}_{x} =  \sum\nolimits_{l \in \mathcal{L}}\sum\nolimits_{k \in \mathcal{K}_{l}}\boldsymbol{W}_{l,k} + \boldsymbol{R}_{0}$ is also approximated as the sample covariance matrix at the BS. Based on $\eqref{FIM_def_2}$, the FIM is derived in Lemma \ref{FIM_lem} as follows.
 \begin{lem}\label{FIM_lem}
 	The FIM $\boldsymbol{J}_{l}$ for estimating $\boldsymbol{\xi}_{l,k}$ is given by
 	\begin{align}
 		\boldsymbol{J}_{l} = \left[\begin{array}{cc}
 			{J}_{\theta_l,\theta_l}&\boldsymbol{J}_{\theta_l,\boldsymbol{\beta}_{l}}\\
 			\boldsymbol{J}_{\theta_l,\boldsymbol{\beta}_{l}}^{T}& \boldsymbol{J}_{\boldsymbol{\beta}_{l},\boldsymbol{\beta}_{l}}
 		\end{array}\right],\label{FIM}
 	\end{align}
 	where 
 	\begin{align}
 		&{J}_{\theta_l,\theta_l}= \frac{2T\left|\beta_l\right|^2}{\sigma_{s}^{2}}\mathrm{tr}\left(\dot{\boldsymbol{E}}_{l}(\theta_l)\boldsymbol{\Phi}_{l}\boldsymbol{G}_{l}\boldsymbol{R}_{x}\boldsymbol{G}_{l}^{H}\boldsymbol{\Phi}_{l}^{H}\dot{\boldsymbol{E}}_{l}^{H}(\theta_l)\right) ,\label{J_theta}\\
 		&\boldsymbol{J}_{\theta_l,\boldsymbol{\beta}_{l}} =\nonumber\\ &\frac{2T}{\sigma_{s}^{2}}\mathrm{Re}\left\{{\beta}_{l}^{*}\mathrm{tr}\left({\boldsymbol{E}}_{l}(\theta_l)\boldsymbol{\Phi}_{l}\boldsymbol{G}_{l}\boldsymbol{R}_{x}\boldsymbol{G}_{l}^{H}\boldsymbol{\Phi}_{l}^{H}\dot{\boldsymbol{E}}_{l}^{H}(\theta_l)\right)[1,j]\right\},\label{J_theta_beta}\\
 		&\boldsymbol{J}_{\boldsymbol{\beta}_{l},\boldsymbol{\beta}_{l}} = \frac{2T}{\sigma_{s}^{2}}\mathrm{tr}\left({\boldsymbol{E}}_{l}(\theta_l)\boldsymbol{\Phi}_{l}\boldsymbol{G}_{l}\boldsymbol{R}_{x}\boldsymbol{G}_{l}^{H}\boldsymbol{\Phi}_{l}^{H}{\boldsymbol{E}}_{l}^{H}(\theta_l)\right)\boldsymbol{I}_{2},\label{J_beta}
 	\end{align}
 	with $\dot{\boldsymbol{E}}_{l}(\theta_l) = \frac{\partial \boldsymbol{E}_{l}}{\partial \theta_l}$ denoting the partial derivative of $\boldsymbol{E}_{l}$ with respect to $\theta_l$. 
 \end{lem}
\begin{IEEEproof}
	See Appendix \ref{FIM_lem_proof}.
\end{IEEEproof}

	Next, based on the FIM $\boldsymbol{J}_{l}$, we are particularly interested in the CRB for each IRS $l$ to estimate $\theta_l$,\footnote{As the complex coefficient $\beta_l$ is affected by both the radar cross section and the round-trip path loss, it is difficult for IRS $l$ to extract target information from $\beta_l$. Therefore, only the CRB for estimating $\theta_l$ is considered.} which is obtained as 
	\begin{align}
		\overline{\mathrm{CRB}}_l({\theta_l}) = \left[\boldsymbol{J}_{l}^{-1}\right]_{1,1}= \left[{J}_{{\theta_l},{\theta_l}} - \boldsymbol{J}_{{\theta_l},\tilde{\boldsymbol{\beta}}_{l}}\boldsymbol{J}_{\tilde{\boldsymbol{\beta}}_{l},\tilde{\boldsymbol{\beta}}_{l}}^{-1}\boldsymbol{J}_{{\theta_l},\tilde{\boldsymbol{\beta}}_{l}}^{T} \right]^{-1}.\label{CRB_theta1}
	\end{align} 
	For $\boldsymbol{E}_{l} = \tilde{\boldsymbol{a}}_{l}(\theta_l)\boldsymbol{a}_{l}^{T}(\theta_l)$, it follows that
	\begin{align}
		&\dot{\boldsymbol{E}}_{l}(\theta_l) = \dot{\tilde{\boldsymbol{a}}}_{l}(\theta_l)\boldsymbol{a}_{l}^{T}(\theta_l)+\tilde{\boldsymbol{a}}_{l}(\theta_l)\dot{\boldsymbol{a}}_{l}^{T}(\theta_l)\nonumber\\
		&=j \frac{2\pi}{\lambda}d_s \cos (\theta_l) \left(\boldsymbol{D}_{N_s}\tilde{\boldsymbol{a}}_{l}(\theta_l)\boldsymbol{a}_{l}^{T}(\theta_l)+\tilde{\boldsymbol{a}}_{l}(\theta_l)\boldsymbol{a}_{l}^{T}(\theta_l)\boldsymbol{D}_{N}\right), \label{E_dot}
	\end{align}
	where $\boldsymbol{D}_{N} = \mathrm{diag}(0,1,\ldots,N-1)$. 
	Based on \eqref{E_dot} and \eqref{CRB_theta1}, we have the following proposition.
	\begin{prop}\label{CRB_lem}
		The CRB for each IRS to estimate $\theta_l$ is given by \eqref{CRB_theta} and \eqref{CRB_theta2} at the top of next page,
		\newcounter{TempEqCnt} 
		\begin{figure*}[h]
				\begin{align}\label{CRB_theta}
						&\overline{\mathrm{CRB}}_l({\theta_l})=\frac{\sigma_{\mathrm{s}}^2}{2 T|\beta_l|^2\left(\mathrm{tr}\left(\dot{\boldsymbol{E}}_{l}(\theta_l)\boldsymbol{\Phi}_{l}\boldsymbol{G}_{l}\boldsymbol{R}_x \boldsymbol{G}_{l}^{H}\boldsymbol{\Phi}_{l}^{H}\dot{\boldsymbol{E}}_{l}^{H}(\theta_l)\right)-\frac{\left|\mathrm{tr}\left({\boldsymbol{E}}_{l}(\theta_l)\boldsymbol{\Phi}_{l}\boldsymbol{G}_{l}\boldsymbol{R}_{x}\boldsymbol{G}_{l}^{H}\boldsymbol{\Phi}_{l}^{H}\dot{\boldsymbol{E}}_{l}^{H}(\theta_l)\right)\right|^2}{\mathrm{tr}\left({\boldsymbol{E}}_{l}(\theta_l)\boldsymbol{\Phi}_{l}\boldsymbol{G}_{l}\boldsymbol{R}_{x}\boldsymbol{G}_{l}^{H}\boldsymbol{\Phi}_{l}^{H}{\boldsymbol{E}}_{l}^{H}(\theta_l)\right)}\right)}\\
						&=\frac{\sigma_{\mathrm{s}}^2 \lambda^2}{8 T|\beta_l|^2\pi^2 d_s^2 \cos^2(\theta_l)\left(\frac{{\left( {{N_s} - 1} \right){N_s}\left( {{N_s} + 1} \right)}}{{12}}{\mathop{\rm tr}\limits} \left( {\boldsymbol{\phi}_{l}^{T}{\boldsymbol{U}_l}\boldsymbol{\phi}_{l}^{*}} \right) + {N_s}{\rm{tr}}\left( {\boldsymbol{\phi}_{l}^{T}{{\boldsymbol{D}}_N}{\boldsymbol{U}_l}{{\boldsymbol{D}}_N}\boldsymbol{\phi}_{l}^*} \right) - \frac{{{N_s}{{\left| {{\mathop{\rm tr}\limits} \left( {{\boldsymbol{\phi}_{l}^{T}}{\boldsymbol{U}_l}{{\boldsymbol{D}}_N}\boldsymbol{\phi}_{l}^*} \right)} \right|}^2}}}{{{\mathop{\rm tr}\limits} \left( {{\boldsymbol{\phi}_{l}^{T}}{\boldsymbol{U}_l}\boldsymbol{\phi}_{l}^*} \right)}}\right)},  \label{CRB_theta2}   
					\end{align}	
			\hrulefill
				\vspace{-4mm}
		\end{figure*}
		where ${\boldsymbol{U}_l} = {{{\boldsymbol{A}}_l}{{\boldsymbol{G}}_{l}}{{\boldsymbol{R}}_x}{\boldsymbol{G}}_{l}^H{\boldsymbol{A}}_l^H}$ and  $\boldsymbol{A}_{l} = \mathrm{diag} (\boldsymbol{a}_l(\theta_l))$. 
	\end{prop}
	 \begin{IEEEproof}
See Appendix \ref{CRB_lem_proof}.
	 \end{IEEEproof}
\begin{rem} 
\textcolor{black}{Note that the CRB in \eqref{CRB_theta} and \eqref{CRB_theta2} is dependent on the value of $\theta_{l}$. In the direction of interest, where a potential target may reside, we can minimize the $\overline{\mathrm{CRB}}_l({\theta_l})$  by jointly optimizing the transmit beamforming at the BS and the reflective beamforming at the IRSs. In certain radar scenarios, e.g., target tracking, beamforming does not necessitate frequent updates when the target is moving slowly \cite{liu2022cramer}. In such instances, employing an estimated or predicted direction is sufficient for beamforming design. Based on these factors, we assume that a rough prior estimation of the target's parameters is available. This assumption enables us to optimize the pertinent CRB with the goal of enhancing real-time estimation accuracy. It is worth noting that such a consideration has been widely adopted in the literature \cite{li2007range}.} 
\end{rem}

Based on the closed-form CRB expression in \eqref{CRB_theta2}, we have the following proposition to reveal the relationship between CRB and the number of sensors $N_s$ at each IRS. 
\begin{prop}\label{CRB_limit_1}
When the number of sensing elements  $N_s$ at each IRS becomes sufficiently large,	the CRB of estimating $\theta_{l}$ for the point target case decreases inversely proportional to the cubic of  $N_s$, i.e.,  $\mathrm{CRB}_{l}(\theta_l) \propto \frac{1}{N_s^{3}}$.	
\end{prop}
	 \begin{IEEEproof}
  See Appendix \ref{CRB_limit_1_proof}.
\end{IEEEproof}
	

\subsection{Extended Target Case}
Next, we consider the extended target case, in which the objective of each IRS $l$ is to estimate  the complete target response matrix $\hat{\boldsymbol{E}}_{l} \in \mathbb{C}^{N_s \times N}$. By vectorizing the received echo signal in \eqref{EchoSigAtIRS} at IRS $l$, we have  
\begin{align}
			\hat{\boldsymbol{y}}_{l} = \mathrm{vec}(\bar{\boldsymbol{Y}}_{l}) = \hat{\boldsymbol{\eta}}_{l} + \tilde{\boldsymbol{z}}_{l},\label{Vec_Y_extended}
\end{align}
where
\begin{align}
\hat{\boldsymbol{\eta}}_{l} = \mathrm{vec} (\hat{\boldsymbol{E}}_{l}\boldsymbol{\Phi}_{l}\boldsymbol{G}_{l}\boldsymbol{X}) = \left((\boldsymbol{\Phi}_{l}\boldsymbol{G}_{l}\boldsymbol{X})^{T} \otimes \boldsymbol{I}_{N_s}\right)\hat{\boldsymbol{h}}_{l}, \label{Extended_target_signal}
\end{align}
and $\hat{\boldsymbol{h}}_{l} = \mathrm{vec}(\hat{\boldsymbol{E}}_{l})$. Based on the definition of FIM in \eqref{FIM_def_2} and the received echo signal model in \eqref{Extended_target_signal}, the FIM for estimating $\hat{\boldsymbol{h}}_{l}$ is given in the following lemma. 
\begin{lem}\label{FIM_extended_lem}
	The FIM $\boldsymbol{F}_{l}$ for estimating $\hat{\boldsymbol{h}}_{l}$ from \eqref{Vec_Y} is given by 
	\begin{align}
		\boldsymbol{F}_{l} = \left[\begin{array}{cc}
			\boldsymbol{F}_{\mathrm{Re}\{\hat{\boldsymbol{h}}_{l}\},\mathrm{Re}\{\hat{\boldsymbol{h}}_{l}\}}&\boldsymbol{F}_{\mathrm{Re}\{\hat{\boldsymbol{h}}_{l}\},\mathrm{Im}\{\hat{\boldsymbol{h}}_{l}\}}\\
			\boldsymbol{F}_{\mathrm{Im}\{\hat{\boldsymbol{h}}_{l}\},\mathrm{Re}\{\hat{\boldsymbol{h}}_{l}\}}& \boldsymbol{F}_{\mathrm{Im}\{\hat{\boldsymbol{h}}_{l}\},\mathrm{Im}\{\hat{\boldsymbol{h}}_{l}\}}
		\end{array}\right], \label{FIM_extended}
	\end{align}
\end{lem}
where 
\begin{align}
		&\boldsymbol{F}_{\mathrm{Re}\{\hat{\boldsymbol{h}}_{l}\},\mathrm{Re}\{\hat{\boldsymbol{h}}_{l}\}} =\boldsymbol{F}_{\mathrm{Im}\{\hat{\boldsymbol{h}}_{l}\},\mathrm{Im}\{\hat{\boldsymbol{h}}_{l}\}}\nonumber\\ &=\frac{2}{\sigma_{{s}}^{2}}\mathrm{Re}\left\{(\boldsymbol{\Phi}_{l}\boldsymbol{G}_{l}\boldsymbol{X})^{*} (\boldsymbol{\Phi}_{l}\boldsymbol{G}_{l}\boldsymbol{X})^{T} \otimes \boldsymbol{I}_{N_s} \right\}, \label{FIM_Imh_Imh}\\
		&\boldsymbol{F}_{\mathrm{Re}\{\hat{\boldsymbol{h}}_{l}\},\mathrm{Im}\{\hat{\boldsymbol{h}}_{l}\}}=-\boldsymbol{F}_{\mathrm{Im}\{\hat{\boldsymbol{h}}_{l}\},\mathrm{Re}\{\hat{\boldsymbol{h}}_{l}\}}\nonumber\\ &=\frac{2}{\sigma_{{s}}^{2}}\mathrm{Re}\left\{(\boldsymbol{\Phi}_{l}\boldsymbol{G}_{l}\boldsymbol{X})^{*} (\boldsymbol{\Phi}_{l}\boldsymbol{G}_{l}\boldsymbol{X})^{T} \otimes \boldsymbol{I}_{N_s} \right\}.\label{FIM_Reh_Imh}
\end{align}
\begin{IEEEproof}
	See Appendix \ref{FIM_extended_lem_proof}.
\end{IEEEproof}
Based on \eqref{FIM_extended}, we have the following proposition. 
\begin{prop}
	\textcolor{black}{The CRB for estimation the target response matrix $\hat{\boldsymbol{E}}_{l}$ by IRS $l$ is given by 
	\begin{align}
		\widetilde{\mathrm{CRB}}_l(\hat{\boldsymbol{E}}_{l})&=\mathrm{tr}(\boldsymbol{F}_{l}^{-1}) =\frac{N_s\sigma_{{s}}^2}{T}\mathrm{tr}\left((\boldsymbol{\Phi}_{l}^{*}\boldsymbol{G}_{l}^{*}\boldsymbol{R}_{x}^{*}\boldsymbol{G}_{l}^{T}\boldsymbol{\Phi}_{l}^{T})^{-1}\right)\nonumber\\
		&=\frac{N_s\sigma_{{s}}^2}{T}\mathrm{tr}\left((\boldsymbol{\Phi}_{l}^{T})^{-1}(\boldsymbol{G}_{l}^{*}\boldsymbol{R}_{x}^{*}\boldsymbol{G}_{l}^{T})^{-1}(\boldsymbol{\Phi}_{l}^{*})^{-1}\right)\nonumber\\
		&=\frac{N_s\sigma_{{s}}^2}{T}\mathrm{tr}\left((\boldsymbol{G}_{l}^{*}\boldsymbol{R}_{x}^{*}\boldsymbol{G}_{l}^{T})^{-1}(\boldsymbol{\Phi}_{l}^{T}\boldsymbol{\Phi}_{l}^{*})^{-1}\right)\nonumber\\
		&=\frac{N_s\sigma_{{s}}^2}{T}\mathrm{tr}\left((\boldsymbol{G}_{l}\boldsymbol{R}_{x}^{H}\boldsymbol{G}_{l}^{H})^{-1}\right). \label{CRB_extend}
	\end{align}}
\end{prop}
\begin{IEEEproof}
This proposition can be verified following a similar procedure as in \cite[(11)]{liu2022cramer}. Therefore, the detailed proof is omitted for brevity. 
\end{IEEEproof}

It is observed from \eqref{CRB_extend} that the $\widetilde{\mathrm{CRB}}_l(\hat{\boldsymbol{E}}_{l})$ only depends on the transmit covariance $\boldsymbol{R}_{x}$ or the transmit beamformers. Besides, the condition $M \geq \mathrm{rank}(\boldsymbol{R}_{x})\geq N \geq \mathrm{rank}(\boldsymbol{G}_{l}) $ must hold to ensure $\widetilde{\mathrm{CRB}}_l(\hat{\boldsymbol{E}}_{l})$ to be bounded, such that the target response matrix $\hat{\boldsymbol{E}}_{l}$ is estimable. \textcolor{black}{As indicated in \eqref{CRB_extend}, the estimated CRB for the extended target includes the channel coefficients $\boldsymbol{G}_{l}$, which can be acquired through appropriate channel estimation techniques \cite{zheng2022survey}. Consequently, we can minimize the CRB for the extended target by optimizing the transmit beamforming at the BS.} Moreover, we have the following proposition to reveal the relationship between CRB and the number of sensors $N_s$ at each IRS.
\begin{prop}\label{CRB_limit_2}
	When the number of sensing elements  $N_s$ at each IRS becomes sufficiently large, the CRB of estimating $\hat{\boldsymbol{E}}_{l}$ for the extended target case increases proportional to $N_s$, i.e., $\widetilde{\mathrm{CRB}}_l(\hat{\boldsymbol{E}}_{l}) \propto {N_s}$.	
\end{prop}
\begin{IEEEproof}
 See Appendix \ref{CRB_limit_2_proof}.	
\end{IEEEproof}

 By comparing Propositions \ref{CRB_limit_1} versus \ref{CRB_limit_2}, it is observed that deploying more reflecting elements at IRS is beneficial to enhance the performance for target DoA estimation (in terms of lower CRB) for the point target case, but leads to higher CRB for the extended target case. This is due to the fact that higher array gain can be exploited for sensing in the former case, but more target  parameters need to be estimated in the latter case. 
	\section{Joint Beamforming for CRB Minimization with Point Targets}
	
	In this section, we jointly optimize the transmit beamforming at the BS and the reflective beamforming at the multiple IRSs to minimize the estimation CRB $\overline{\mathrm{CRB}}_l({\theta_l})$ in \eqref{CRB_theta2} for the point target case, subject to the maximum transmit power constraint at the BS and the SINR requirements at individual CUs. To ensure the fair optimization of sensing performance, we particularly  minimize the maximum CRB among the $L$ IRSs. As such, for Type-I and Type-II CU receivers, the SINR-constrained max-CRB minimization problems are formulated as (P1-I) and (P1-II), respectively. 
		\begin{subequations}
		\begin{eqnarray}
				&\!\!\!\!\!\!\!\!\!\!\!\!\!\!\!\!\!\!(\text{P1-I}):&\nonumber\\
				&\!\!\!\!\!\!\!\!\!\!{\mathop {\min}\limits_{\{\boldsymbol{W}_{l,k}\},\{\boldsymbol{\Phi}_{l}\},\boldsymbol{R}_{0} } } \!\!\!\!&\mathop{\max} \limits_{l \in \mathcal{L}}  \quad \overline{\mathrm{CRB}}_l({\theta_l}) \nonumber\\
			&\!\!\!\!\!\!\!\!\!\!\text{s.t.} \!\!\!\!\!\!\!& \gamma_{l,k}^{(\text{I})} (\{\boldsymbol{W}_{l,k}\},\{\boldsymbol{\Phi}_{l}\},\boldsymbol{R}_{0}) \geq \Gamma_{l,k}, \nonumber\\
			&&\forall l \in \mathcal{L}, \forall k \in \mathcal{K}_{l}  \label{P1_I_cons1}\\
			&&\mathrm{tr}\left(\sum\limits_{l \in \mathcal{L}}\sum\limits_{k \in \mathcal{K}_{l}}\boldsymbol{W}_{l,k} + \boldsymbol{R}_{0}\right) \leq P_{\text{sum}}\label{P1_I_cons2}\\	
			&&\boldsymbol{R}_{0} \succeq \boldsymbol{0} \label{P1_I_cons3}\\
			&&\boldsymbol{W}_{l,k} \succeq \boldsymbol{0}, \forall l \in \mathcal{L}, \forall k \in \mathcal{K}_{l} \label{P1_I_cons4}\\
			&&{\rm rank}\left(\boldsymbol{W}_{l,k}\right) = 1, \forall l \in \mathcal{L}, \forall k \in \mathcal{K}_{l} \label{P1_I_cons5}\\
			&&|\phi_{l,n}| = 1, \forall n \in \mathcal{N}, \forall l \in \mathcal{L}.  \label{P1_I_cons6}
		\end{eqnarray}
	\end{subequations}
	\begin{subequations}
		\begin{eqnarray}
				&\!\!\!\!\!\!\!\!\!\!(\text{P1-II}):&\nonumber\\
				&\!\!\!\!\!\!\!\!\!\!{\mathop {\min}\limits_{\{\boldsymbol{W}_{l,k}\},\{\boldsymbol{\Phi}_{l}\},\boldsymbol{R}_{0} } }  \!\!\!\!&\mathop{\max} \limits_{k \in \mathcal{L}}  \quad \overline{\mathrm{CRB}}_l({\theta_l}) \nonumber\\
			&\!\!\!\!\!\!\!\!\!\!\text{s.t.}  \!\!\!\!& \gamma_{l,k}^{(\text{II})} (\{\boldsymbol{W}_{l,k}\},\{\boldsymbol{\Phi}_{l}\},\boldsymbol{R}_{0}) \geq \Gamma_{l,k}, \nonumber\\
			&&\forall l \in \mathcal{L}, \forall k \in \mathcal{K}_{l} \label{P1_II_cons1}\\
			&&\eqref{P1_I_cons2}-\eqref{P1_I_cons6}. \nonumber
		\end{eqnarray}
	\end{subequations}
In problems (P1-I) and (P1-II), \eqref{P1_I_cons1}, \eqref{P1_I_cons2} and \eqref{P1_I_cons6} denote the SINR constraints at the CUs, the transmit power constraint at the BS, and the unit-modular constraints at the IRSs, respectively.
	
Notice that problems (P1-I) and (P1-II) are both non-convex. This is due to the fact that their objective functions are non-convex and the SINR constraints in \eqref{P1_I_cons1} and \eqref{P1_II_cons1}, the unit-modulus constraints in \eqref{P1_I_cons6}, and the rank-one constraints in \eqref{P1_I_cons5} are all non-convex. To tackle the non-convexity, we propose alternating optimization-based algorithms to solve them, in which the transmit beamforming $\{\boldsymbol{W}_{l,k}\}$/$\boldsymbol{R}_{0}$ at the BS and the reflective beamforming $\{\boldsymbol{\Phi}_{l}\}$ at  IRSs are optimized in an alternating manner. \textcolor{black}{Note that problems (P1-I) and (P1-II) are similar, with the only difference in the SINR $\gamma_{l,k}^{(\text{I})}$ in \eqref{sinr_I} for Type-I receivers versus $\gamma_{l,k}^{(\text{II})}$ in \eqref{sinr_II} for Type-II receivers. Due to their similar structure, the proposed algorithm for solving problem (P1-I) can be directly applied to solve (P1-II) by replacing $\gamma_{l,k}^{(\text{II})}$ with $\gamma_{l,k}^{(\text{I})}$. In particular, we first focus on solving problem (P1-I), and omit the details for solving problem (P1-II).} In the following, we obtain the optimal transmit beamformers $\{\boldsymbol{W}_{l,k}\}$ and $\boldsymbol{R}_{0}$ for problem (P1-I) under given reflective beamformers $\{\boldsymbol{\Phi}_{l}\}$ in Section IV-A, and then optimize $\{\boldsymbol{\Phi}_{l}\}$ for (P1-I) under given $\{\boldsymbol{W}_{l,k}\}$ and $\{\boldsymbol{R}_{0}\}$ in Section IV-B.
	
	\subsection{Optimal Transmit Beamforming to (P1-I) with Given Reflective Beamforming}
	
	First, we aim to optimize the transmit beamforming $\{\boldsymbol{W}_{l,k}\}$ and $\boldsymbol{R}_{0}$ at the BS with given $\{\boldsymbol{\Phi}_{l}\}$, for which the optimization problem is expressed as
	\begin{subequations}
		\begin{eqnarray}
				(\text{P2}):&{\mathop {\min}\limits_{\{\boldsymbol{W}_{l,k}\},\boldsymbol{R}_{0} } }   &  \mathop{\max} \limits_{l \in \mathcal{L}}\quad\overline{\mathrm{CRB}}_l({\theta_l})  \nonumber\\
			&\text{s.t.} & \eqref{P1_I_cons1}-\eqref{P1_I_cons3},\eqref{P1_I_cons4}, \text{and }\eqref{P1_I_cons5}.\nonumber
		\end{eqnarray}
	\end{subequations}
	In (P2), we use the CRB formula in \eqref{CRB_theta2} for transmit beamforming optimization. Note that problem (P2) is still non-convex due to the non-convexity of the objective function, the SINR constraints in \eqref{P1_I_cons1}, and the rank constraints in \eqref{P1_I_cons5}. In the following, we use the SDR technique to find the optimal solution. 
	
	First, we define $\tilde{\boldsymbol{G}}_{l,k} = \boldsymbol{G}_{l}^{H}\boldsymbol{\Phi}_{l}^{H}\boldsymbol{h}_{l,k}\boldsymbol{h}_{l,k}^{H}\boldsymbol{\Phi}_{l}\boldsymbol{G}_{l}$, and equivalently express the SINR constraints in \eqref{P1_I_cons1} as
		\begin{align} \label{SINR_cons_eq}
			&\frac{1}{\Gamma_{l,k}}\mathrm{tr}\left(\tilde{\boldsymbol{G}}_{l,k}\boldsymbol{W}_{l,k} \right) \geq \nonumber\\
			&\sum\limits_{l' \in \mathcal{L} }{\sum\limits_{k' \in \mathcal{L'} \backslash k} \mathrm{tr}\left({\tilde{\boldsymbol{G}}_{l,k}\boldsymbol{W}_{l',k'}}\right) + \mathrm{tr}\left(\tilde{\boldsymbol{G}}_{l,k} \boldsymbol{R}_{0}\right)+ \sigma^2}. 
		\end{align}
Furthermore, it is observed from \eqref{CRB_theta2} that $\overline{\mathrm{CRB}}_l({\theta_l})= \frac{\kappa}{f_{1,l} (\{\boldsymbol{W}_{l,k}\},\boldsymbol{R}_{0}) - f_{2,l} (\{\boldsymbol{W}_{l,k}\},\boldsymbol{R}_{0})}$, where $\kappa = \frac{\sigma_{\mathrm{s}}^2 \lambda^2}{8 T|\beta_l|^2\pi^2 d_s^2 \cos^2(\theta_l)}$ is a constant, and 
\begin{small}
		\begin{align}
		&f_{1,l} (\{\boldsymbol{W}_{l,k}\},\boldsymbol{R}_{0}) = \nonumber\\
		&\frac{{\left( {{N_s} - 1} \right){N_s}\left( {{N_s} + 1} \right)}}{{12}}{\mathop{\rm tr}\limits} \left( {\boldsymbol{\phi}_{l}^{T}{\boldsymbol{B}_{l}\left(\sum\limits_{l \in \mathcal{L} }\sum\limits_{k \in \mathcal{K}_{l}}\boldsymbol{W}_{l,k} + \boldsymbol{R}_{0}\right)\boldsymbol{B}_{l}^{H}}\boldsymbol{\phi}_{l}^*} \right) \nonumber\\
		&  \quad + {N_s}{\rm{tr}}\left( {\boldsymbol{\phi}_{l}^{T}{{\boldsymbol{D}}_N}{\boldsymbol{B}_{l}\left(\sum\limits_{l \in \mathcal{L} }\sum\limits_{k \in \mathcal{K}_{l}}\boldsymbol{W}_{l,k} + \boldsymbol{R}_{0}\right)\boldsymbol{B}_{l}^{H}}{{\boldsymbol{D}}_N}\boldsymbol{\phi}_{l}^*} \right),\\
		&f_{2,l} (\{\boldsymbol{W}_{l,k}\},\boldsymbol{R}_{0}) = \nonumber\\
		&\frac{{{N_s}{{\left| {{\mathop{\rm tr}} \left( {\boldsymbol{\phi}_{l}^{T}{\boldsymbol{B}_{l}\left(\sum\limits_{l \in \mathcal{L} }\sum\limits_{k \in \mathcal{K}_{l}}\boldsymbol{W}_{l,k} + \boldsymbol{R}_{0}\right)\boldsymbol{B}_{l}^{H}}{{\boldsymbol{D}}_N}\boldsymbol{\phi}_{l}^*} \right)} \right|}^2}}}{{{\mathop{\rm tr}} \left( {\boldsymbol{\phi}_{l}^{T}{\boldsymbol{B}_{l}\left(\sum\limits_{l \in \mathcal{L} }\sum\limits_{k \in \mathcal{K}_{l}}\boldsymbol{W}_{l,k} + \boldsymbol{R}_{0}\right)\boldsymbol{B}_{l}^{H}}\boldsymbol{\phi}_{l}^*} \right)}}, 
	\end{align}
\end{small}with $\boldsymbol{B}_{l} = {{\boldsymbol{A}}_l}{{\boldsymbol{G}}_{l}}$. As such, problem (P2) is can be equivalently solved via solving the following problem: 
	\begin{subequations}
		\begin{eqnarray}
				(\text{P2.1}):&{\mathop {\max}\limits_{\{\boldsymbol{W}_{l,k}\}, \boldsymbol{R}_{0}} }  \!\!\! & \mathop{\min} \limits_{l \in \mathcal{L}}  f_{1,l} (\{\boldsymbol{W}_{l,k}\},\boldsymbol{R}_{0}) - f_{2,l} (\{\boldsymbol{W}_{l,k}\},\boldsymbol{R}_{0})\nonumber\\
			&\text{s.t.}\!\!\!& \eqref{P1_I_cons2}-\eqref{P1_I_cons5}, \text{and } \eqref{SINR_cons_eq}.  \nonumber    
		\end{eqnarray}
	\end{subequations}
	
	To solve problem (P2.1), we introduce the auxiliary variables $\tilde{\nu}_{1}$ and $\{\tilde{\nu}_{2,l}\}$. Accordingly, problem (P2.1) is equivalently reformulated as 
		\begin{subequations}
		\begin{eqnarray}
			\!\!\!\!\!\!\!\!\!\!\!\!\!\!\!\!	(\text{P2.2}):& \!\!\!\!\!\!\!\!{\mathop {\max}\limits_{\substack{\{\boldsymbol{W}_{l,k}\}, \boldsymbol{R}_{0},\\ \tilde{\nu}_{1}, \{\tilde{\nu}_{2,l}\}}} } \!\!\!\!& \tilde{\nu}_{1}  \nonumber\\
			&\!\!\!\!\!\!\!\!\text{s.t.}\!\!\!\!& f_{1,l} (\{\boldsymbol{W}_{l,k}\},\boldsymbol{R}_{0})- \tilde{\nu}_{2,l}\geq {\tilde{\nu}_{1} },  \forall l \in \mathcal{L}\label{P2_2_cons1}\\
			&&\tilde{\nu}_{1} \geq 0 \label{P2_2_cons2}\\
			&& \tilde{\nu}_{2,l} \geq f_{2,l} (\{\boldsymbol{W}_{l,k}\},\boldsymbol{R}_{0}), \forall l \in \mathcal{L} \label{P2_2_cons3}\\
			&&\eqref{P1_I_cons2}-\eqref{P1_I_cons5},\text{and } \eqref{SINR_cons_eq}.  \nonumber      
		\end{eqnarray}
	\end{subequations}
To deal with the non-convex constraints in \eqref{P2_2_cons3}, we transform them into a set of linear matrix inequality (LMI) constraints base on the Schur’s complement, which is given in \eqref{LMI_cons} at the top of the next page.	
\begin{figure*}[h]
		\begin{small}
	\begin{align}
		\left[\begin{array}{cc}
			{\mathop{\rm tr}\limits} \left( {\boldsymbol{\phi}_{l}^{T}{\boldsymbol{B}_{l}\left(\sum\limits_{l \in \mathcal{L} }\sum\limits_{k \in \mathcal{K}_{l}}\boldsymbol{W}_{l,k} + \boldsymbol{R}_{0}\right)\boldsymbol{B}_{l}^{H}}\boldsymbol{\phi}_{l}^*} \right) &  \sqrt{N_s}{ {{\mathop{\rm tr}\limits} \left( {\boldsymbol{\phi}_{l}^{T}{\boldsymbol{B}_{l}\left(\sum\limits_{l \in \mathcal{L} }\sum\limits_{k \in \mathcal{K}_{l}}\boldsymbol{W}_{l,k} + \boldsymbol{R}_{0}\right)\boldsymbol{B}_{l}^{H}}{{\boldsymbol{D}}_N}\boldsymbol{\phi}_{l}^*} \right)} }\\ 
			\sqrt{N_s}{ {{\mathop{\rm tr}\limits} \left( {\boldsymbol{\phi}_{l}^{T}{{\boldsymbol{D}}_N}{\boldsymbol{B}_{l}\left(\sum\limits_{l \in \mathcal{L} }\sum\limits_{k \in \mathcal{K}_{l}}\boldsymbol{W}_{l,k}^{H} + \boldsymbol{R}_{0}^{H}\right)\boldsymbol{B}_{l}^{H}}\boldsymbol{\phi}_{l}^*} \right)} }  &    \tilde{\nu}_{2,l}
		\end{array}\right] \succeq 0, \forall l \in \mathcal{L}. \label{LMI_cons}
	\end{align}  
\end{small}	
	\vspace{-4mm}
\end{figure*}   
As a result, problem (P2.2) is equivalent to the following convex problem:
	\begin{subequations}
		\begin{eqnarray}
				(\text{P2.3}):& {\mathop {\max}\limits_{\substack{\{\boldsymbol{W}_{l,k}\}, \boldsymbol{R}_{0},\\ \tilde{\nu}_{1}, \{\tilde{\nu}_{2,l}\}}} } &\!\!\! \tilde{\nu}_{1}  \nonumber\\
			&\text{s.t.}& \!\!\!\eqref{P1_I_cons2}-\eqref{P1_I_cons5}, \eqref{SINR_cons_eq}, \eqref{P2_2_cons1}, \eqref{P2_2_cons2}, \text{and } \eqref{LMI_cons}.  \nonumber      
		\end{eqnarray}
	\end{subequations}
	
Furthermore, we remove the rank constraints in \eqref{P1_I_cons5}, and accordingly obtain the relaxed version of (P2.3) as (SDR2.3).	
		Note that (SDR2.3) is a convex semi-definite program (SDP), which can be optimally solved by convex solvers such as CVX \cite{grant2014cvx}. Let $\{\overline{\boldsymbol{W}}_{l,k}\}$ and $\overline{\boldsymbol{R}_{0}}$ denote the obtained optimal solution to (SDR2.3). As the obtained optimal solution $\{\overline{\boldsymbol{W}}_{l,k}\}$ may not be of rank-one, we construct the optimal rank-one solutions of $\{\overline{\boldsymbol{W}}_{l,k}^{\star}\}$ and the corresponding $\overline{\boldsymbol{R}}_{0}^{\star}$ to problem (P2.3) and (P2) by using the following proposition. 
	\begin{prop}\label{prop1}
		Based on the obtained optimal solution $\{\overline{\boldsymbol{W}}_{l,k}\}$ and $\overline{\boldsymbol{R}_{0}}$ to (SDR2.3), the optimal solution to problem (P2.3) and (P2) is given by 
			\begin{align}
			\overline{\boldsymbol{W}}_{l,k}^{\star} &= \overline{\boldsymbol{w}}_{l,k}\overline{\boldsymbol{w}}_{l,k}^{H}, \forall l\in \mathcal{L}, k \in \mathcal{K}_{l},  \label{prop1.2}
		\end{align}
		with $\overline{\boldsymbol{w}}_{l,k} = \left(\tilde{\boldsymbol{h}}_{l,k}^{H}\overline{\boldsymbol{W}}_{l,k}\tilde{\boldsymbol{h}}_{l,k}\right)^{-\frac{1}{2}}\overline{\boldsymbol{W}}_{l,k}\tilde{\boldsymbol{h}}_{l,k}$, where $\tilde{\boldsymbol{h}}_{l,k}^{H} = \boldsymbol{h}_{l,k}^{H}\boldsymbol{\Phi}_{l}\boldsymbol{G}_{l}$, and the corresponding optimal solution of $\overline{\boldsymbol{R}}_{0}^{\star}$ is given by 
		\begin{align}		
			\overline{\boldsymbol{R}}_{0}^{\star} &= \overline{\boldsymbol{R}}_{0}+\sum\limits_{l \in \mathcal{L} }\sum\limits_{k \in \mathcal{K}_{l}}\overline{\boldsymbol{W}}_{l,k}-\sum\limits_{l \in \mathcal{L} } \sum\limits_{k \in \mathcal{K}_{l}}\overline{\boldsymbol{W}}_{l,k}^{\star}.\label{prop1.3}
		\end{align}
	\end{prop}
	\begin{IEEEproof}
 See Appendix \ref{prop1_proof}.		
	\end{IEEEproof}
	
	It is shown in Proposition \ref{prop1} that the solutions $\{\overline{\boldsymbol{W}}_{l,k}^{\star}\}$ and $\overline{\boldsymbol{R}}_0^{\star}$ in \eqref{prop1.2} and \eqref{prop1.3} are actually optimal for problem (SDR2.2). Therefore, the SDR is tight between (P2.2) and (SDR2.2), and thus we obtain the optimal solution to (P2). 

	\subsection{Reflective Beamforming Optimization for (P1-I) with Given Transmit Beamforming}
	
	Next, we optimize the reflective beamforming $\{\boldsymbol{\Phi}_{l}\}$ or $\{\boldsymbol{\phi}_{l}\}$ for problem (P1-I) with given $\{\boldsymbol{w}_{l,k}\}$ and $\boldsymbol{R}_{0}$. In this case, the optimization problem is expressed as 
	\begin{subequations}
		\begin{eqnarray}
				(\text{P3}):&{\mathop {\min}\limits_{\{\boldsymbol{\phi}_{l}\}} } &  \mathop{\max} \limits_{l \in \mathcal{L}}\quad \overline{\mathrm{CRB}}_l({\theta_l}) \nonumber\\ 
				&\text{s.t.}&  \eqref{P1_I_cons1},\eqref{P1_I_cons6}, \nonumber
		\end{eqnarray}
	\end{subequations}
in which we use the CRB formulas in \eqref{CRB_theta2} for reflective beamforming optimization.  Notice that problem (P3) can be equivalently decomposed into $L$ subproblems given by  
	\begin{subequations}
		\begin{eqnarray}
				(\text{P3.1.$l$}):&{\mathop {\text{max}}\limits_{\boldsymbol{\phi}_{l}} } & \frac{{\left( {{N_s} - 1} \right){N_s}\left( {{N_s} + 1} \right)}}{{12}}{\mathop{\rm tr}\limits} \left( {\phi_{l}^{T}\boldsymbol{U}_{l}\boldsymbol{\phi}_{l}^*} \right)  \nonumber\\
				&&+ {N_s}{\rm{tr}}\left( {\boldsymbol{\phi }_{l}^{T}{{\boldsymbol{D}}_N}\boldsymbol{U}_{l}{{\boldsymbol{D}}_N}\boldsymbol{\phi}_{l}^*} \right) \nonumber\\
				& &\quad - \frac{{{N_s}{{\left| {{\mathop{\rm tr}\limits} \left( {\boldsymbol{\phi}_{l}^{T}\boldsymbol{U}_{l}{{\boldsymbol{D}}_N}\boldsymbol{\phi}_{l}^*} \right)} \right|}^2}}}{{{\mathop{\rm tr}\limits} \left( {\boldsymbol{\phi}_{l}^{T}\boldsymbol{U}_{l}\boldsymbol{\phi}_{l}^*} \right)}}  \nonumber\\
			&\text{s.t.}&\eqref{P1_I_cons1},\eqref{P1_I_cons6}. \nonumber
		\end{eqnarray}
	\end{subequations}
	Problem (P3.1.$l$) is still non-convex due to the non-convexity of the objective function, the SINR constraints in  \eqref{P1_I_cons1}, and the unit-modulus constraints in \eqref{P1_I_cons6}. 	

	In the following, we use the SDR technique to handle problem (P3.1.$l$). Towards this end, we define $\boldsymbol{H}_{l,k} =  \mathrm{diag}(\boldsymbol{h}_{l,k}^{H})$ and $\boldsymbol{\Theta}_{l} = \boldsymbol{\phi}_{l}^{*}\boldsymbol{\phi}_{l}^{T}$, where $\boldsymbol{\Theta}_{l} \succeq \boldsymbol{0}$ and $\mathrm{rank} \left(\boldsymbol{\Theta}_{l}\right) = 1$, $\forall l \in \mathcal{L}, k\in\mathcal{K}_{l}$. 
	Then, the SINR at CU $k$ in the coverage area of IRS $l$ becomes \eqref{SINR_eq2} at the top of this page.
	\begin{figure*}
			\begin{align}\label{SINR_eq2}
			&\tilde{\gamma}_{l,k}^{(\text{I})} (\{\boldsymbol{\Theta}_{l}\}) =  \frac{{{\mathrm{tr}{\left( \boldsymbol{H}_{l,k}\boldsymbol{G}_{l}\boldsymbol{W}_{l,k}\boldsymbol{G}_{l}^{H}\boldsymbol{H}_{l,k}^{H}\boldsymbol{\Theta}_{l}  \right)}}}}{{\sum\limits_{l' \in \mathcal{L} }\sum\limits_{k' \in \mathcal{K}_{l'} \backslash k}\!\!\!\! \!{\mathrm{tr}{{\left(\boldsymbol{H}_{l,k}\boldsymbol{G}_{l}\boldsymbol{W}_{l',k'}\boldsymbol{G}_{l}^{H}\boldsymbol{H}_{l,k}^{H}\boldsymbol{\Theta}_{l}\right)}}} \!+\!\mathrm{tr}\!\left(\boldsymbol{H}_{l,k}\boldsymbol{G}_{l} \boldsymbol{R}_{0}\boldsymbol{G}_{l}^{H} \boldsymbol{H}_{l,k}^{H}\boldsymbol{\Theta}_{l}\right) \!+\! \sigma^2}}. 
		\end{align}
		\hrulefill
  \vspace{-3mm}
	\end{figure*}
	
By substituting \eqref{SINR_eq2} into the constraints in \eqref{P1_I_cons1} and skipping the rank-one constraints, problem (P3.1.$l$) is relaxed as 
	\begin{subequations}
		\begin{eqnarray}
				&\!\!\!\!\!\!\!\!\!\!\!\!(\text{SDR3.1.$l$}): &\nonumber\\
				 &\!\!\!\!\!\!\!\!\!\!\!\!\!\!\!\!\!\!\!\!\!\!\!\!\!\!\!\!\!\!{\mathop {\text{max}}\limits_{\boldsymbol{\Theta}_{l}} } &\!\!\!\!\!\!\!\! \!\!\!\!\!\!\frac{{\left( {{N_s} - 1} \right){N_s}\left( {{N_s} + 1} \right)}}{{12}}{\mathop{\rm tr}\limits} \left( {{\boldsymbol{U}_l}\boldsymbol{\Theta}_{l}} \right)  \nonumber\\
				&&\!\!\!\!\!\!\!\!\!\!\!\!\!\!+ {N_s}{\rm{tr}}\left( {{{\boldsymbol{D}}_N}{\boldsymbol{U}_l}{{\boldsymbol{D}}_N}\boldsymbol{\Theta}_{l}} \right) - \frac{{{N_s}{{\left| {{\mathop{\rm tr}\limits} \left( {{\boldsymbol{U}_l}{{\boldsymbol{D}}_N}\boldsymbol{\Theta}_{l}} \right)} \right|}^2}}}{{{\mathop{\rm tr}\limits} \left( {{\boldsymbol{U}_l}\boldsymbol{\Theta}_{l}} \right)}}  \nonumber\\
			&\!\!\!\!\!\!\!\!\!\!\!\!\!\!\!\!\!\!\!\!\!\!\!\!\!\!\text{s.t.}&\!\!\!\!\!\!\!\!\!\!\!\!\!\!\!\!\!\!\!\!\!\!\!\!\sum\limits_{l' \in \mathcal{L} }\sum\limits_{k' \in \mathcal{K}_{l'} \backslash k} {\mathrm{tr}{{\left(\boldsymbol{H}_{l,k}\boldsymbol{G}_{l}\boldsymbol{W}_{l',k'}\boldsymbol{G}_{l}^{H}\boldsymbol{H}_{l,k}^{H}\boldsymbol{\Theta}_{l}\right)}}} \nonumber\\
			&&\!\!\!\!\!\!\!\!\!\!\!\!\!\!\!\!\!\!\!\!\!\!\!\! +\mathrm{tr}\left(\boldsymbol{H}_{l,k}\boldsymbol{G}_{l} \boldsymbol{R}_{0}\boldsymbol{G}_{l}^{H} \boldsymbol{H}_{l,k}^{H}\boldsymbol{\Theta}_{l}\right) + \sigma^2 \nonumber\\ 
			&&\!\!\!\!\!\!\!\!\!\!\!\!\!\!\!\!\!\!\!\!\!\!\!\!-\frac{1}{\Gamma_{l,k} }{{{\mathrm{tr}{\left( \boldsymbol{H}_{l,k}\boldsymbol{G}_{l}\boldsymbol{W}_{l,k}\boldsymbol{G}_{l}^{H}\boldsymbol{H}_{l,k}^{H}\boldsymbol{\Theta}_{l} \right)}}}} \leq 0, \forall k \in \mathcal{K}_{l}\label{P8_cons1}\\
			&&\!\!\!\!\!\!\!\!\!\!\!\!\!\!\!\!\!\!\!\!\!\!\!\![\boldsymbol{\Theta}_{l}]_{n,n} = 1, \forall n \in \mathcal{N}\label{P8_cons2}\\
			&&\!\!\!\!\!\!\!\!\!\!\!\!\!\!\!\!\!\!\!\!\!\!\!\!\boldsymbol{\Theta}_{l} \succeq \boldsymbol{0}. \label{P8_cons3}
		\end{eqnarray}
	\end{subequations}
Furthermore, by introducing an auxiliary variable $\tau_k$, problem (SDR3.1.$l$) is equivalently re-expressed as 
	\begin{subequations}
		\begin{eqnarray}
				&\!\!\!\!\!\!\!\!\!\!\!\!\!\!\!\!\!\!(\text{SDR3.2.$l$}):&\nonumber\\
				&\!\!\!\!\!\!\!\!\!\!\!\!\!\!\!\!\!\!\!\!\!\!\!\!{\mathop {\text{max}}\limits_{\boldsymbol{\Theta}_{l}, \tau_l} } &\!\!\!\!\!\!\!\!\!\!\!\!\!\!\!\!\!\!\frac{{\left( {{N_s} - 1} \right){N_s}\left( {{N_s} + 1} \right)}}{{12}}{\mathop{\rm tr}\limits} \left( {{\boldsymbol{U}_l}\boldsymbol{\Theta}_{l}} \right)\nonumber\\
				&&\!\!\!\!\!\!\!\!\!\!\!\!\!\!\!\!\!\!+ {N_s}{\rm{tr}}\left( {{{\boldsymbol{D}}_N}{\boldsymbol{U}_l}{{\boldsymbol{D}}_N}\boldsymbol{\Theta}_{l}} \right) - \tau_l  \nonumber\\
			 &\!\!\!\!\!\!\!\!\!\!\!\!\!\!\!\!\!\!\!\!\!\!\!\!\text{s.t.} &
			 \!\!\!\!\!\!	\!\!\!\!\!\!\!\!\!\!\!\!\left[\begin{array}{cc}
			 		\tau_k &  \!\!\!\!\!\!\sqrt{N_s}{{\mathop{\rm tr}\limits} \left( {{\boldsymbol{U}_l}{{\boldsymbol{D}}_N}\boldsymbol{\Theta}_{l}} \right)}\\
			 		\sqrt{N_s}{{\mathop{\rm tr}\limits} \left( {\boldsymbol{\Theta}_{l}^{H}{{\boldsymbol{D}}_N}{\boldsymbol{U}_l^{H}}} \right)}\!\!\!\!\!\! \!\!\!\!\!\! & \mathrm{tr} \left( {{\boldsymbol{U}_l}\boldsymbol{\Theta}_{l}} \right) 
			 	\end{array}\right] \succeq 0
\label{relax_cons}\\
			 &&\!\!\!\!\!\!	\!\!\!\!\!\!\!\!\!\!\!\!\eqref{P8_cons1}-\eqref{P8_cons3}. \nonumber
		\end{eqnarray}
	\end{subequations}
    Here, the constraints in \eqref{relax_cons} is obtained from $\tau_l \geq \frac{{{N_s}{{\left| {{\mathop{\rm tr}\limits} \left( {{\boldsymbol{U}_l}{{\boldsymbol{D}}_N}\boldsymbol{\Theta}_{l}} \right)} \right|}^2}}}{{{\mathop{\rm tr}\limits} \left( {{\boldsymbol{U}_l}\boldsymbol{\Theta}_{l}} \right)}}$ by using Schur's complement.
	Note that problem (SDR3.2.$l$) is a convex SDP which can be optimally solved by CVX. Let $\boldsymbol{\Theta}_{l}^{\star}$ denote the obtained optimal solution to (SDR3.2.$l$). 
	
	Notice that the obtained solution $\boldsymbol{\Theta}_{l}^{\star}$ may not be of rank-one, and as a result, it may not be the optimal solution to problem (P3.1.$l$) or (P3). To tackle this issue, we further use the Gaussian randomization to construct a high-quality solution to problem (P3.1.$l$) based on the obtained $\left\{\boldsymbol{\Theta}_{l}^{\star}\right\}$. Specifically, we first generate a number of randomly vectors ${\boldsymbol{r}}_{l} \sim \mathcal{CN}\left(\boldsymbol{0},\boldsymbol{I}_N\right)$, and then construct a number of rank-one solutions as 
	\begin{align}\label{GR_construct}
		\boldsymbol{\phi}_{l} = e^{j\mathrm{arg}\left\{\left(\boldsymbol{\Theta}_{l}^{\star}\right)^{\frac{1}{2}}\boldsymbol{r}_{l}\right\}}.
	\end{align}	
Then, we find the desirable solution of $\boldsymbol{\phi}_{l}$ that minimizes $\overline{\mathrm{CRB}}_l({\theta_l})$ among all random generated $\boldsymbol{\phi}_{l}$'s.  As a result, problem (P3) is finally solved. 
	 
	\textcolor{black}{In summary, the alternating optimization-based algorithm for solving problem (P1-I) is implemented by solving problems (P2) and (P3) alternately. Notice that problem (P2) is optimally solved and the solution to (P3) would lead to a decreasing max-CRB with a sufficient number of Gaussian randomizations. As such, the alternating optimization-based algorithm leads to monotonically non-increasing max-CRB values over iterations, and thus is ensured to converge.} Furthermore, the alternating optimization-based algorithm can be performed in a distributed way in practice. \textcolor{black}{In particular, in each iteration, the BS can first optimize the transmit beamforming by solving problem (P2), and then each IRS can optimize its own reflective beamforming by solving problem (P3.1.$l$) in a distributed manner. The proposed algorithm is thus efficient in practical implementation.}
	\section{Joint Beamforming for CRB Minimization with Extended Targets}
	In this section, we jointly optimize the transmit beamforming $\{\boldsymbol{W}_{l,k}\}$ and $\boldsymbol{R}_{0}$ at the BS and the reflective beamforming $\{\boldsymbol{\Phi}_{l}\}$ at the IRSs to minimize the maximum estimation CRB among the IRSs, subject to the maximum transmit power constraint at the BS and the SINR requirements at individual CUs. For Type-I and Type-II CU receivers, the SINR-constrained max-CRB minimization problems are formulated as (P4-I) and (P4-II) in the following, respectively.
		\begin{subequations}
			\begin{eqnarray}
					(\text{P4-I}):&{\mathop {\min}\limits_{\{\boldsymbol{W}_{l,k}\},\{\boldsymbol{\Phi}_{l}\},\boldsymbol{R}_{0} } }    & \mathop{\max} \limits_{l \in \mathcal{L}}  \quad \widetilde{\mathrm{CRB}}_l(\hat{\boldsymbol{E}}_{l}) \nonumber\\
				 &\text{s.t.}& \eqref{P1_I_cons1}-\eqref{P1_I_cons6}.\nonumber
			\end{eqnarray}
			\begin{eqnarray}
					(\text{P4-II}):&{\mathop {\min}\limits_{\{\boldsymbol{W}_{l,k}\},\{\boldsymbol{\Phi}_{l}\},\boldsymbol{R}_{0} } }   & \mathop{\max} \limits_{l \in \mathcal{L}}  \quad  \widetilde{\mathrm{CRB}}_l(\hat{\boldsymbol{E}}_{l}) \nonumber\\
				 &\text{s.t.}& \eqref{P1_II_cons1},\eqref{P1_I_cons2}-\eqref{P1_I_cons6}.\nonumber
			\end{eqnarray}
		\end{subequations}
Problems (P4-I) and (P4-II) are both non-convex. This is due to the fact that their objective functions are non-convex, the SINR constraints in \eqref{P1_I_cons1} and \eqref{P1_II_cons1}, the unit-modulus constraints in \eqref{P1_I_cons6}, the rank-one constraints in \eqref{P1_I_cons5} are all non-convex. To tackle the non-convexity, we propose an alternating optimization-based algorithm to solve them by optimizing $\{\boldsymbol{W}_{l,k}\}/\boldsymbol{R}_{0}$ and $\{\boldsymbol{\Phi}_{l}\}$ alternately. \textcolor{black}{As problem (P4-II) has a similar structure as problem (P4-I), in the following, we only focus on solving problem (P4-I) by omitting the details for solving (P4-II).}
	\subsection{Optimal Transmit Beamforming to (P4-I) with Given Reflective Beamforming}
	First, we aim to optimize the transmit beamforming $\{\boldsymbol{W}_{l,k}\}$ and dedicated signal covariance $\boldsymbol{R}_{0}$ with given $\{\boldsymbol{\Phi}_{l}\}$. In this case, the optimization problem is formulated as
	\begin{subequations}
		\begin{align}
				&(\text{P5}):&{\mathop {\min}\limits_{\{\boldsymbol{W}_{l,k}\},\boldsymbol{R}_{0} } }   &\mathop{\max} \limits_{l \in \mathcal{L}}\quad  \widetilde{\mathrm{CRB}}_l(\hat{\boldsymbol{E}}_{l})  \nonumber\\
			& &\text{s.t.}\quad & \eqref{P1_I_cons1}-\eqref{P1_I_cons5}.\nonumber
		\end{align}
	\end{subequations}
	By introducing an auxiliary variable $\tilde{u}$, problem (P5) is equivalent to the following problem. 
	\begin{subequations}
		\begin{eqnarray}
				&\!\!\!\!\!\!\!\!\!\!\!\!\!(\text{P5.1}):&\nonumber\\
				&\!\!\!\!\!\!\!\!\!\!\!\!\!{\mathop {\min}\limits_{\{\boldsymbol{W}_{l,k}\}, \boldsymbol{R}_{0}, \tilde{u}}}  &  \tilde{u}  \nonumber\\
			&\!\!\!\!\!\!\!\!\!\!\!\!\! \text{s.t.} &\!\!\!\!\!\!\!\!\!\! \frac{N_s\sigma_{{s}}^2}{T}\mathrm{tr}\left(\!\!\!\left(\boldsymbol{G}_{l}\left(\sum\limits_{l \in \mathcal{L} }\sum\limits_{k \in \mathcal{K}_{l} }\boldsymbol{W}_{l,k} + \boldsymbol{R}_{0}\right)^{\!\!\! H}\!\!\!\!\boldsymbol{G}_{l}^{H}\right)^{\!\!\! -1}\right)\nonumber\\
			&&\!\!\!\!\!\!\!\!\!\! \leq \tilde{u}, \forall l \in \mathcal{L} \label{P3_3_cons1}\\
			&&\!\!\!\!\!\!\!\!\!\!\eqref{P1_I_cons2}-\eqref{P1_I_cons5}, \text{and }\eqref{SINR_cons_eq}.  \nonumber      
		\end{eqnarray}
	\end{subequations}
	We then use the SDR technique to handle problem (P5.1). By removing  the rank-one constraints in \eqref{P1_I_cons5}, we obtain the relaxed version of (P5.1) as (SDR5.1). Note that problem (SDR5.1) is a convex SDP problem that can be solved by CVX.  Let $\{\widetilde{\boldsymbol{W}}_{l,k}\}$ and $\widetilde{\boldsymbol{R}_{0}}$ denote the obtained optimal solution to problem (SDR5.1). The optimal rank-one solution $\{\widetilde{\boldsymbol{W}}_{l,k}^{\star}\}$ and the corresponding $\widetilde{\boldsymbol{R}}_{0}^{\star}$ to problems (P5) and (P5.1) can be constructed via the following proposition.
	\begin{prop}
		Based on the obtained optimal solution $\{\widetilde{\boldsymbol{W}}_{l,k}\}$ and $\widetilde{\boldsymbol{R}_{0}}$ to (SDR5.1), the optimal  solution of $\{\boldsymbol{W}_{l,k}\}$ to problem (P5) and (P5.1) is given by 
		\begin{align}
			\widetilde{\boldsymbol{W}}_{l,k}^{\star} = \widetilde{\boldsymbol{w}}_{l,k}\widetilde{\boldsymbol{w}}_{l,k}^{H}, \forall k\in \mathcal{L},
		\end{align}
		with $\widetilde{\boldsymbol{w}}_{l,k} = \left(\tilde{\boldsymbol{h}}_{l,k}^{H}\widetilde{\boldsymbol{W}}_{l,k}\tilde{\boldsymbol{h}}_{l,k}\right)^{-\frac{1}{2}}\widetilde{\boldsymbol{W}}_{l,k}\tilde{\boldsymbol{h}}_{l,k}$, and the corresponding ${\boldsymbol{R}}_{0}^{\star}$ is given by 
		\begin{align}		
		\widetilde{\boldsymbol{R}}_{0}^{\star} = \widetilde{\boldsymbol{R}}_{0}+\sum\limits_{l \in \mathcal{L} } \sum\limits_{k \in \mathcal{K}_{l}}\widetilde{\boldsymbol{W}}_{l,k}-\sum\limits_{l \in \mathcal{L} } \sum\limits_{k \in \mathcal{K}_{l}}\widetilde{\boldsymbol{W}}_{l,k}^{\star}.
		\end{align}
	\end{prop}
	\begin{IEEEproof}
		The proof is similar to that of Proposition \ref{prop1}, and thus is omitted.
	\end{IEEEproof}

	\subsection{Reflective Beamforming Optimization for (P4-I) with Given Transmit Beamforming}
	Next, we optimize the reflecting coefficients $\{\boldsymbol{\phi}_{l}\}$ in problem  (P4-I) with given $\{\boldsymbol{W}_{l,k}\}$ and $\boldsymbol{R}_{0}$. Note that the objective function is independent of $\{\boldsymbol{\phi}_{l}\}$. As a result,  any $\{\boldsymbol{\phi}_{l}\}$ that guarantee the SINR constraints in \eqref{P1_I_cons1} is a feasible solution to problem (P4-\text{I}). \textcolor{black}{As the reflective beamformers $\{\boldsymbol{\phi}_{l}\}$ may affect the optimization of  $\{\boldsymbol{W}_{l,k}\}$ and $\boldsymbol{R}_{0}$ by changing the achieved maximum SINR, we propose to optimize $\{\boldsymbol{\phi}_{l}\}$ to maximize the minimum SINR at CUs for expanding the feasible region of problem (P5). Note that the SINR of each CU $k \in \mathcal{K}_{l}$ only depends on its associated IRS $l$. Therefore, we optimize the reflective beamforming at each IRS $l$ to maximize the minimum SINR at the covered CUs $\{k\}$ where $k \in \mathcal{K}_{l}$. The corresponding optimization problem becomes (P6.$l$) at the top of the next page.}
	\begin{figure*}\textcolor{black}{
			\begin{subequations}
			\begin{eqnarray}
					(\text{P6.$l$}):&{\mathop {\max}\limits_{\boldsymbol{\Phi}_{l}} } &  \mathop{\min}_{k\in\mathcal{K}_{l}} \frac{{{ \boldsymbol{h}_{l,k}^{H}\boldsymbol{\Phi}_{l}\boldsymbol{G}_{l}\boldsymbol{W}_{l,k}\boldsymbol{G}_{l}^{H}\boldsymbol{\Phi}_{l}^{H}\boldsymbol{h}_{l,k} }}}{\sum\limits_{l' \in \mathcal{L} } {\sum\limits_{k' \in \mathcal{K}_{l'} \backslash k}\!\!\!\!\!\! {{\boldsymbol{h}_{l,k}^{H}\boldsymbol{\Phi}_{l}\boldsymbol{G}_{l}\boldsymbol{W}_{l',k'}\boldsymbol{G}_{l}^{H}\boldsymbol{\Phi}_{l}^{H} \boldsymbol{h}_{l,k} }} \!\!+\!\!\boldsymbol{h}_{l,k}^{H}\boldsymbol{\Phi}_{l}\boldsymbol{G}_{l} \boldsymbol{R}_{0}\boldsymbol{G}_{l}^{H} \boldsymbol{\Phi}_{l}^{H}\boldsymbol{h}_{l,k}\!\!+\!\! \sigma^2}}  \nonumber\\
				&\text{s.t.} & \eqref{P1_I_cons6}.\nonumber
			\end{eqnarray}
		\end{subequations}}
		\vspace{-1cm}
	\end{figure*}

Furthermore, by replacing the objective function in problem (P6) by $\tilde{\gamma}_{l,k}^{\text{I}}$ in \eqref{SINR_eq2} and skipping the rank-one constraints on $\boldsymbol{\Theta}_{l}$, we have a relaxed version of (P6) as problem (SDR6) at the top of next page.
\begin{figure*}\textcolor{black}{
\begin{subequations}
		\begin{eqnarray}
				(\text{SDR6.$l$}):&{\mathop {\max}\limits_{\boldsymbol{\Theta}_{l}} }   & \mathop{\min}_{k\in\mathcal{K}_{l}} {\frac{{{\mathrm{tr}{\left( \boldsymbol{H}_{l,k}\boldsymbol{G}_{l}\boldsymbol{W}_{l,k}\boldsymbol{G}_{l}^{H}\boldsymbol{H}_{l,k}^{H}\boldsymbol{\Theta}_{l}  \right)}}}}{{\sum\limits_{l' \in \mathcal{L} } \sum\limits_{k' \in \mathcal{K}_{l'} \backslash k}\!\! {\mathrm{tr}{{\left(\boldsymbol{H}_{l,k}\boldsymbol{G}_{l}\boldsymbol{W}_{l',k'}\boldsymbol{G}_{l}^{H}\boldsymbol{H}_{l,k}^{H}\boldsymbol{\Theta}_{l}\right)}}} +\mathrm{tr}\left(\boldsymbol{H}_{l,k}\boldsymbol{G}_{l} \boldsymbol{R}_{0}\boldsymbol{G}_{l}^{H} \boldsymbol{H}_{l,k}^{H}\boldsymbol{\Theta}_{l}\right)+ \sigma^2}}} \nonumber  \\ 
				&\text{s.t.}& [\boldsymbol{\Theta}_{l}]_{n,n} = 1, \forall n \in \mathcal{N} \\
				&&\boldsymbol{\Theta}_{l} \succeq \boldsymbol{0}.  
			\end{eqnarray}
		\end{subequations}}
			\hrulefill
    \vspace{-15pt}
\end{figure*}

Problem (SDR6.$l$) is still a non-convex problem as the objective function is non-convex. To tackle this issue, we define \textcolor{black}{
		\begin{align}
			&\boldsymbol{T}_{1,l,k} = \boldsymbol{H}_{l,k}\boldsymbol{G}_{l}\boldsymbol{w}_{l,k}\boldsymbol{w}_{l,k}^{H}\boldsymbol{G}_{l}^{H}\boldsymbol{H}_{l,k}^{H},\\ &\boldsymbol{T}_{2,l,k} = \nonumber\\
			&\sum\limits_{l' \in \mathcal{L} } \sum\limits_{k' \in \mathcal{K}_{l'} \backslash k}\!\!\!\!\!\!\boldsymbol{H}_{l,k}\boldsymbol{G}_{l}\boldsymbol{w}_{l',k'}\boldsymbol{w}_{l',k'}^{H}\boldsymbol{G}_{l}^{H}\boldsymbol{H}_{l,k}^{H} \!+\!\boldsymbol{H}_{l,k}\boldsymbol{G}_{l} \boldsymbol{R}_{0}\boldsymbol{G}_{l}^{H} \boldsymbol{H}_{l,k}^{H}.
		\end{align}
		Then, by introducing an auxiliary variable $\tilde{w}$, problem (SDR6.$l$) is equivalent to the following optimization problem:
		\begin{subequations}
			\begin{eqnarray}
					&(\text{SDR6.$l$.1}):&\nonumber\\
					&{\mathop {\max}\limits_{\boldsymbol{\Theta}_{l}, \tilde{w}} }   & \tilde{w} \nonumber  \\ 
					&\text{s.t.}& \frac{{{\mathrm{tr}{\left(\boldsymbol{T}_{1,l,k}\boldsymbol{\Theta}_{l}  \right)}}}}{{{\mathrm{tr}{{\left(\boldsymbol{T}_{2,l,k}\boldsymbol{\Theta}_{l}\right)}}} + \sigma^2}} \geq \tilde{w}, \forall k \in \mathcal{K}_{l}\\
					&&[\boldsymbol{\Theta}_{l}]_{n,n} = 1, \forall n \in \mathcal{N} \\
					&&\boldsymbol{\Theta}_{l} \succeq \boldsymbol{0}.     
				\end{eqnarray}
			\end{subequations}
Let $\boldsymbol{\Theta}_{l}^{*}$ denote the optimal solution to problem (SDR6.$l$.1) that is also optimal to problem (SDR6.$l$). We employ a bisection method to obtain $\boldsymbol{\Theta}_{l}^{*}$. Specifically, we first define the search range of the objective value to problem (\text{SDR6.$l$.1}) as $\tilde{w} \in [w_{\text{min}},w_{\text{max}}]$ and a tolerance $\epsilon$. In each iteration step $q$, let $\tilde{w}^{(q)} = \frac{w_{\text{min}}+w_{\text{max}}}{2}$ and then we need to check the following feasibility problem: 
\begin{subequations}
	\begin{eqnarray}
			&(\text{SDR6.$l$.2}):&\nonumber\\
			&{\text{Find}}  & \boldsymbol{\Theta}_{l} \nonumber  \\ 
			&\text{s.t.}& \frac{1}{\tilde{w}^{(q)}}{{\mathrm{tr}{\left(\boldsymbol{T}_{1,l,k}\boldsymbol{\Theta}_{l}  \right)}}}-{{{\mathrm{tr}{{\left(\boldsymbol{T}_{2,l,k}\boldsymbol{\Theta}_{l}\right)}}} - \sigma^2}} \nonumber\\
			&&\geq 0, \forall k \in \mathcal{K}_{l}\\
			&&[\boldsymbol{\Theta}_{l}]_{n,n} = 1, \forall n \in \mathcal{N} \\
			&&\boldsymbol{\Theta}_{l} \succeq \boldsymbol{0}.  
		\end{eqnarray}
	\end{subequations}
If problem (\text{SDR6.$l$.2}) is feasible, then we denote $\boldsymbol{\Theta}_{l}^{(q)}$ as the solution to the problem and accordingly set $w_{\text{min}} = \tilde{w}^{(q)}$ and $\boldsymbol{\Theta}_{l}^{*}=\boldsymbol{\Theta}_{l}^{(q)}$. Otherwise, we set $w_{\text{max}} = \tilde{w}^{(q)}$. We perform the above step iteratively until the condition $w_{\text{max}}-w_{\text{min}}\leq \epsilon$ satisfied. Finally, we obtain the optimal solution $\boldsymbol{\Theta}_{l}^{*}$ to problems (\text{SDR6.$l$}) and (\text{SDR6.$l$.1}).} Now, it remains to find a high-quality solution to (P6.$l$). As the obtained solution $\boldsymbol{\Theta}_{l}^{*}$ may not be of rank-one, it may not be the optimal solution to problem (P6.$l$). To overcome this case, the Gaussian randomization method is used to construct a high-quality rank-one solution to problems (P6.$l$). Similar to \eqref{GR_construct}, we first generate a number of random vectors ${\boldsymbol{r}}_{l} \sim \mathcal{CN}\left(\boldsymbol{0},\boldsymbol{I}_N\right)$, and construct a number of rank-one candidates as $\boldsymbol{\phi}_{l} = e^{j\mathrm{arg}\left\{\left(\boldsymbol{\Theta}_{l}^{*}\right)^{\frac{1}{2}}\boldsymbol{r}_{l}\right\}}$. 
Then, we find the desirable solution of $\boldsymbol{\phi}_{l}$ that maximizes the objective function in problem (P6.$l$) among all randomly generated $\boldsymbol{\phi}_{l}$'s.  As a result, problem (P6.$l$) is finally solved. 

	\textcolor{black}{In summary, the alternating optimization-based algorithm for solving (P4-I) is implemented by solving problems (P5) and (P6.$l$) for all $l\in\mathcal{L}$ alternately.} Similarly as for (P1-I), the alternating optimization-based algorithm leads to monotonically non-increasing max-CRB values for problem (P4-I) over iterations, and thus is ensured to converge. Furthermore, the alternating optimization-based algorithm can also be performed in a distributed way in practice.

	\section{Numerical Results}
	
\begin{figure}[tbp]
		\setlength{\abovecaptionskip}{-0pt}
		\setlength{\belowcaptionskip}{-10pt}
		\centering
		\includegraphics[width= 0.41\textwidth]{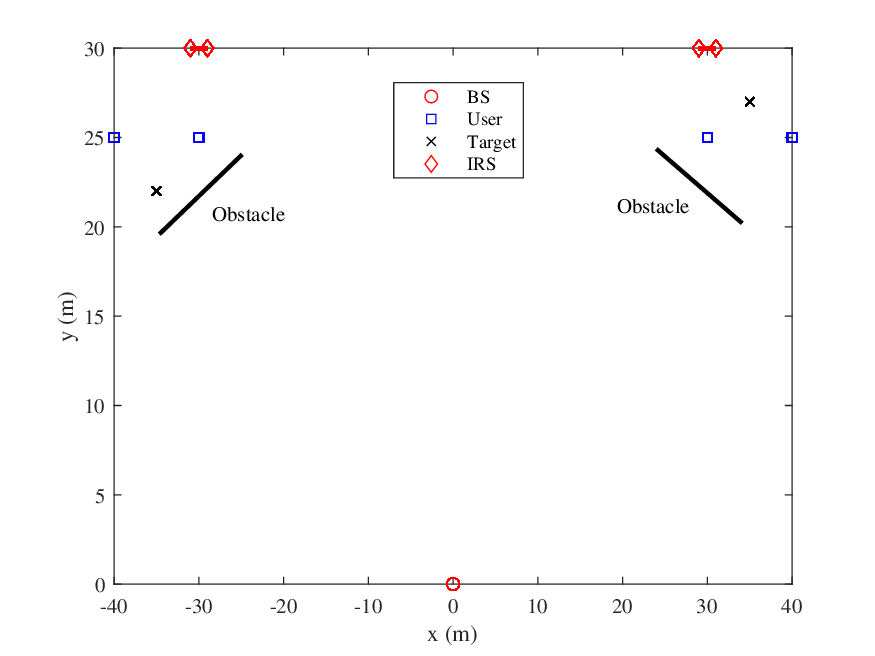}
		\DeclareGraphicsExtensions.
		\caption{\color{black}The location topology.}
		\label{NetworkTopology}
	\end{figure}

	This section provides numerical results to validate the effectiveness of our proposed designs. In the simulation, we set the carrier frequency as $6$ GHz and the bandwidth as $1$ MHz. We adopt the Rician fading channel model with the L-factor being $5$ dB for wireless channels between the BS and the IRSs and those between the IRSs and the CUs. We also set the noise power spectrum density as $-174$ dBm/Hz, the SINR constraints to be $\Gamma_{l,k} = \Gamma, \forall l\in \mathcal{L}, \forall  k \in \mathcal{L}$, and the radar dwell time as $T = 100$.
	\textcolor{black}{In particular, we consider a scenario with one BS and two IRSs, each covering two CUs as shown in Fig. \ref{NetworkTopology}. The BS and two IRSs are located at $(0,0)$ meters~(m),  $(-30,30)$ m, and $(30,30)$ m, respectively. Their associated CUs are located at $(-40,25)$ m, $(-30,25)$ m, and $(30,25)$ m, $(40,25)$ m, and targets are located at $(-35,22)$ m and $(35,27)$ m, respectively.}  For comparison, we consider the following benchmark schemes: 
	
	\begin{itemize}
		\item \textbf{Transmit beamforming (BF) only}: The IRSs implement random reflection coefficients. Accordingly, we only optimize the transmit beamforming at the BS, e.g., by solving problems (P2) for the point target case and by solving problem (P5) for the extended target case, when the Type-I CU receivers are implemented.  		
	\item	\textbf{ZF-BF}: First, we design the transmit information beamformers based on the ZF principle. \textcolor{black}{In particular, by defining $\tilde{\boldsymbol{H}} = [\tilde{\boldsymbol{h}}_{1,1},\cdots,\tilde{\boldsymbol{h}}_{l,K},\cdots,\tilde{\boldsymbol{h}}_{L,1},\cdots,\tilde{\boldsymbol{h}}_{L,K}]$} and   $\tilde{\boldsymbol{W}}^{\text{ZF}} = \tilde{\boldsymbol{H}}\left(\tilde{\boldsymbol{H}}^{H} \tilde{\boldsymbol{H}}\right)^{-1}$, we set the ZF beamformer for each CU $k$ as $\boldsymbol{w}_{l,k}^{\text{ZF}} =  \sqrt{p_{l,k}}\frac{\tilde{\boldsymbol{w}}_{l,k}^{\text{ZF}}}{\left\|\tilde{\boldsymbol{w}}_{l,k}^{\text{ZF}}\right\|}$, where $\tilde{\boldsymbol{w}}_{l,k}^{\text{ZF}}$ denotes the $2(l-1)+k$-th column of $\tilde{\boldsymbol{W}}^{\text{ZF}}$ and $p_{l,k}$ is the transmit power for CU $k$. Next, we optimize the transmit  power $\{p_{l,k}\}$, the sensing beamformers $\boldsymbol{R}_0$, and the reflective beamformers $\{\boldsymbol{\Phi}_k\}$ by alternating optimization, similarly as in Sections IV and V for point and extended targets, respectively. 
	\item 	\textcolor{black}{\textbf{Sensing only}: This scheme corresponds to the proposed design without SINR constraints for communications (only sensing task is considered), by equivalently setting  $\Gamma_{l,k} = 0, \forall k\in\mathcal L$, in problems (P1-I), (P1-II), (P4-I), and (P4-II) for the four considered cases, respectively. This thus  serves as performance upper bound for our proposed design, in terms of the lower bound of the estimation CRB.}
	\end{itemize}
\begin{figure}[tbp]
	\setlength{\abovecaptionskip}{-0pt}
	\setlength{\belowcaptionskip}{-10pt}
	\centering
	\includegraphics[width= 0.41\textwidth]{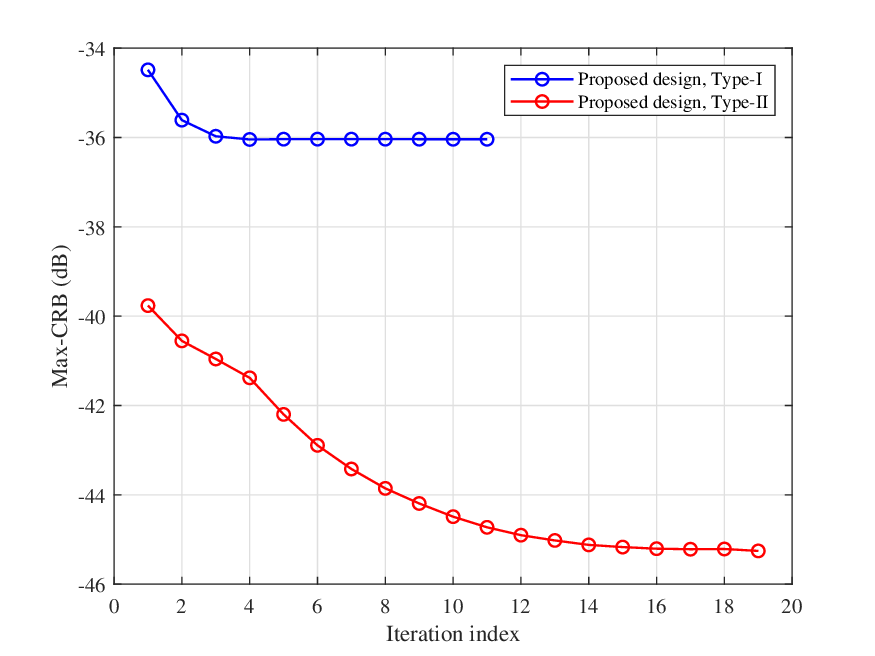}
	\DeclareGraphicsExtensions.
	\caption{\color{black}The achieved max-CRB versus the iteration index for point target case with $\Gamma = 10$dB and $M=N=N_{s}=8$.}
	\label{ConvergencePointTatget}
\end{figure}

\textcolor{black}{Fig. \ref{ConvergencePointTatget} verifies the convergence of the proposed algorithm for the point target case. It is observed that for both two types of CU receivers, the max-CRB performance gradually decreases as the iteration index increases. Consequently, the convergence of the proposed algorithm is guaranteed.}

\begin{figure}[tbp]
	\setlength{\abovecaptionskip}{-0pt}
	\setlength{\belowcaptionskip}{-10pt}
	\centering
	\includegraphics[width= 0.41\textwidth]{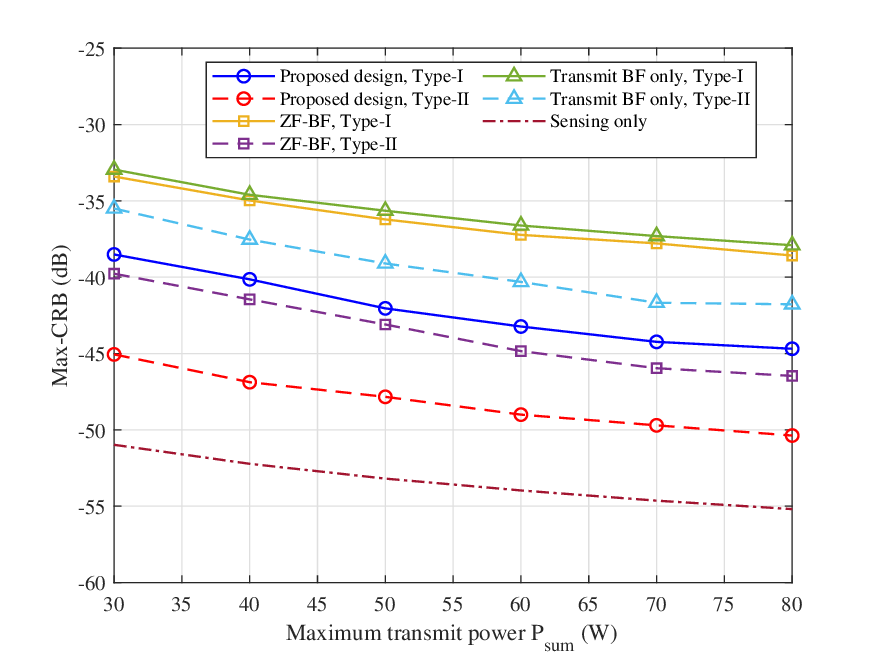}
	\DeclareGraphicsExtensions.
	\caption{\color{black}The achieved max-CRB versus the maximum transmit power $P_{\text{sum}}$ for point target case with $\Gamma = 10$ dB and  $M=N=N_s=8$.}
	\label{CRBVsPower}
\end{figure}

Fig.~\ref{CRBVsPower} shows the achieved max-CRB versus the maximum transmit power $P_{\text{sum}}$ at the BS for the point target case, where we set $\Gamma = 10$ dB. It is observed that our proposed design outperforms other benchmark schemes over the whole regime of $P_{\text{sum}}$. It is also observed that the proposed design with Type-II receivers achieves lower CRB than that with Type-I receivers. This shows the benefits of sensing-signal-interference cancellation. In particular, it is shown in simulations that for problem (P1-II) with Type-II receivers, the obtained $\boldsymbol{R}_0$ is always non-zero for facilitating sensing without interfering with CU receivers; while for problem (P1-I) with Type-I receivers, the obtained $\boldsymbol{R}_0$ is generally zero for avoiding the interference towards CU receivers.

	\begin{figure}[tbp]
		\setlength{\abovecaptionskip}{-0pt}
		\setlength{\belowcaptionskip}{-10pt}
		\centering
		\includegraphics[width= 0.41\textwidth]{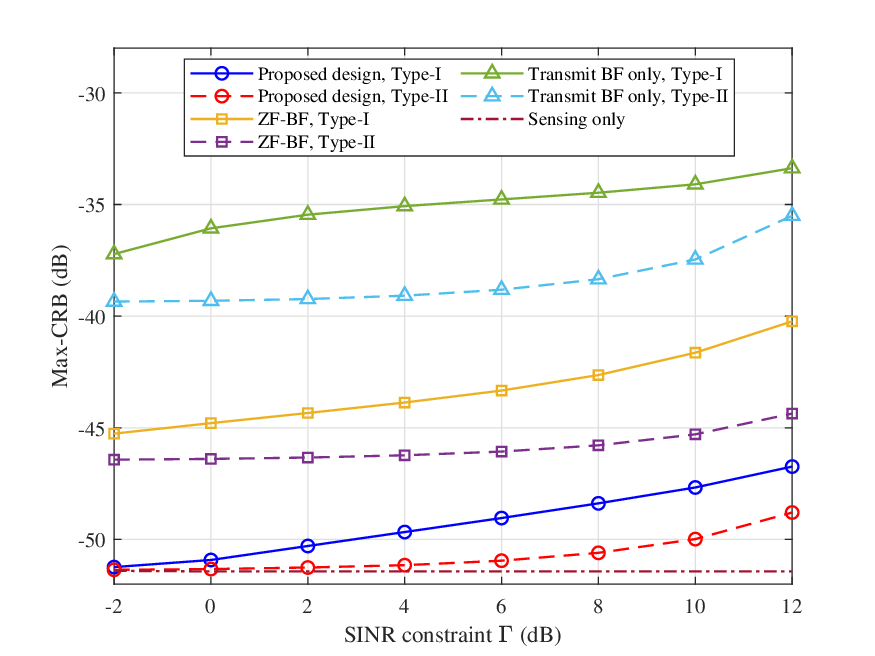}
		\DeclareGraphicsExtensions.
		\caption{\color{black}The achieved max-CRB versus the SINR constraint for point target case with $P_{\text{sum}} = 40$ W and $M=N=N_s=8$.}
		\label{CRBVsSINR_PointTarget}
	\end{figure}
Fig.~\ref{CRBVsSINR_PointTarget} shows the achieved max-CRB versus the SINR constraint $\Gamma$ at each CU for the point target case. It is observed that in the low SINR regime, the performance achieved by our proposed design with both types of CU receivers is close to that by the sensing only scheme. As the SINR constraint increases, the performance gap between the proposed design with Type-II CU receivers and that with Type-I CU receivers is observed to increase. This is due to the fact that when the SINR constraint becomes large, the BS needs to allocate more transmit power to the information signals, which may suffer from more severe interference from dedicated sensing signals when Type-I CU receivers are employed. This shows that the sensing-interference cancellation of Type-II receivers are particularly appealing when both communication and sensing requirements become stringent. 


%
\begin{figure}[tbp]
	\setlength{\abovecaptionskip}{-0pt}
	\setlength{\belowcaptionskip}{-10pt}
	\centering
	\includegraphics[width= 0.41\textwidth]{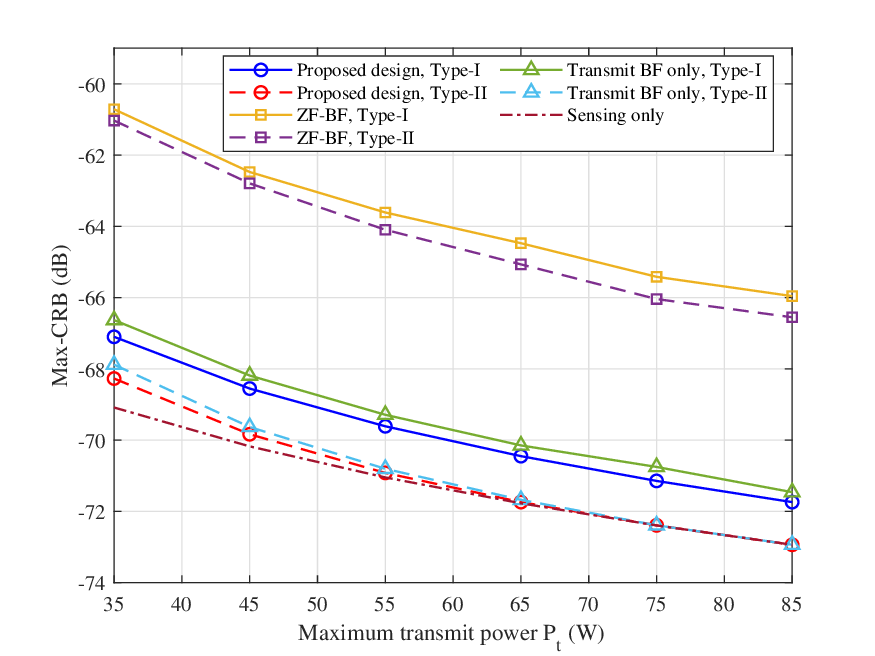}
	\DeclareGraphicsExtensions.
	\caption{\color{black}The achieved max-CRB versus the maximum transmit power $P_{\text{sum}}$ for extended target case with $\Gamma = 10$ dB and $M=N=N_s=8$.}
	\label{CRBVsPower_ExtendedTarget}
\end{figure}
Fig.~\ref{CRBVsPower_ExtendedTarget} shows the achieved max-CRB versus the maximum transmit power $P_{\text{sum}}$ at the BS for extended target case, where we set $\Gamma = 10$ dB. It is observed that our proposed design performs most close to the performance upper bound by the sensing only scheme, and outperforms the transmit beamforming only scheme and ZF beamforming scheme especially when the maximum transmit power $P_{\text{sum}}$ is low. As the maximum transmit power $P_{\text{sum}}$ increases, the CRB performances by our proposed design and the transmit beamforming only scheme with type-II receivers gradually approach the performance upper bound. This is due to the fact that the CRB performance mainly depends on the BS transmit beamforming optimization when the maximum transmit power is high. It is also observed that the proposed design with Type-II receivers achieves lower CRB than that with Type-I receivers. This shows that performing the sensing-interference cancellation of Type-II receivers is also important for the extended target case.

\begin{figure}[tbp]
	\setlength{\abovecaptionskip}{-0pt}
	\setlength{\belowcaptionskip}{-10pt}
	\centering
	\includegraphics[width= 0.41\textwidth]{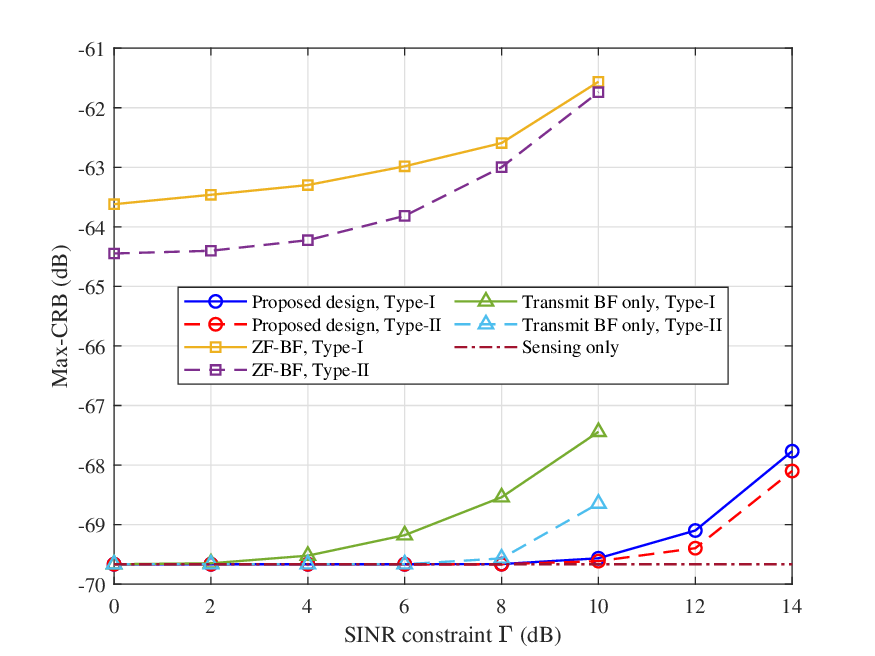}
	\DeclareGraphicsExtensions.
	\caption{\color{black}The minimum CRB versus the SINR requirements for extended target case with $P_{\text{sum}} = 40$ W and $M=N=N_s=8$.}
	\label{CRBVsSINR_ExtendedTarget}
\end{figure}
Fig.~\ref{CRBVsSINR_ExtendedTarget} shows the achieved max-CRB versus the SINR constraint $\Gamma$ at each CU for the extended target case. It is observed that in the low SINR regime, the performance achieved by our proposed design with both two types of CU receivers is close to that by the sensing only scheme. This phenomenon is similar to the point target case. It is also observed that the ZF beamforming scheme performs worst among all schemes. This is different from the observation for the point target case, which indicates the importance of transmit beamforming optimization for minimizing the max-CRB in the extended target case. 

\section{Conclusion}
This paper studied a multi-IRS-enabled ISAC system, in which multiple semi-passive IRSs are deployed to provide ISAC services at separate areas. We derived the closed-form CRBs for each IRS to estimate target parameters based on the ISAC signals, by considering two cases with point and extended targets, respectively. These derived CRBs provides insights on the multi-semi-passive-IRS-enabled sensing. Furthermore, we proposed two efficient joint transmit and reflective beamforming designs to minimize the maximum CRB at all IRSs, while ensuring the communication requirements, by considering two different types of CU receivers. Numerical results showed the effectiveness of our proposed designs, as compared to various benchmarks without such joint optimization. It was also shown that the sensing signal interference cancellation at Type-II receivers is essential in further enhancing the ISAC performance, especially for the point target case.
	
\begin{appendices}
	\section{Proof of Lemma \ref{FIM_lem}}\label{FIM_lem_proof}
	According to \eqref{FIM_def_2}, we have the following the partial derivations:
	\begin{align}
	\frac{\partial \tilde{\boldsymbol{\eta}}}{\partial \theta_{l}}	&= \beta_{l}\mathrm{vec}\left(\dot{\boldsymbol{E}}_{l}(\theta_l)\boldsymbol{\Phi}_{l}\boldsymbol{G}_{l}\boldsymbol{X}\right),\\
	\frac{\partial \tilde{\boldsymbol{\eta}}}{\partial \boldsymbol{\beta}_{l}} &= [1,j] \otimes \mathrm{vec}\left(\boldsymbol{E}_{l}\boldsymbol{\Phi}_{l}\boldsymbol{G}_{l}\boldsymbol{X}\right),
	\end{align}
	where $\dot{\boldsymbol{E}}_{l}(\theta_l) = \frac{\partial \boldsymbol{E}_{l}}{\partial \theta_l}$ denotes the partial derivative of $\boldsymbol{E}_{l}$ with respect to $\theta_l$. Consequently, the elements of FIM in \eqref{FIM_def_2} are given by
	\begin{align}
	 	&{J}_{\theta_l,\theta_l}= \frac{2}{\sigma_{s}^{2}}\mathrm{Re}\left\{\left(\beta_{l}\mathrm{vec}\left(\dot{\boldsymbol{E}}_{l}(\theta_l)\boldsymbol{\Phi}_{l}\boldsymbol{G}_{l}\boldsymbol{X}\right)\right)^H\right.\nonumber\\
	 	&\left.\cdot \beta_{l}\mathrm{vec}\left(\dot{\boldsymbol{E}}_{l}(\theta_l)\boldsymbol{\Phi}_{l}\boldsymbol{G}_{l}\boldsymbol{X}\right)\right\}\nonumber\\
	 	&=\frac{2|\beta_{l}|^2}{\sigma_{s}^{2}}\mathrm{Re}\left\{\mathrm{tr}\left(\left(\dot{\boldsymbol{E}}_{l}(\theta_l)\boldsymbol{\Phi}_{l}\boldsymbol{G}_{l}\boldsymbol{X}\right)^H\dot{\boldsymbol{E}}_{l}(\theta_l)\boldsymbol{\Phi}_{l}\boldsymbol{G}_{l}\boldsymbol{X}\right)\right\}\nonumber\\
	 	&=\frac{2T\left|\beta_l\right|^2}{\sigma_{s}^{2}}\mathrm{tr}\left(\dot{\boldsymbol{E}}_{l}(\theta_l)\boldsymbol{\Phi}_{l}\boldsymbol{G}_{l}\boldsymbol{R}_{x}\boldsymbol{G}_{l}^{H}\boldsymbol{\Phi}_{l}^{H}\dot{\boldsymbol{E}}_{l}^{H}(\theta_l)\right),\\
	&\boldsymbol{J}_{\theta_l,\boldsymbol{\beta}_{l}} =  \frac{2}{\sigma_{s}^{2}} \mathrm{Re}\left\{\left(\beta_{l}\mathrm{vec}\left(\dot{\boldsymbol{E}}_{l}(\theta_l)\boldsymbol{\Phi}_{l}\boldsymbol{G}_{l}\boldsymbol{X}\right)\right)^H  [1,j] \right.\nonumber\\
	&\left.\otimes \mathrm{vec}\left(\boldsymbol{E}_{l}\boldsymbol{\Phi}_{l}\boldsymbol{G}_{l}\boldsymbol{X}\right) \right\} \nonumber\\ &=\frac{2T}{\sigma_{s}^{2}}\mathrm{Re}\left\{{\beta}_{l}^{*}\mathrm{tr}\left({\boldsymbol{E}}_{l}(\theta_l)\boldsymbol{\Phi}_{l}\boldsymbol{G}_{l}\boldsymbol{R}_{x}\boldsymbol{G}_{l}^{H}\boldsymbol{\Phi}_{l}^{H}\dot{\boldsymbol{E}}_{l}^{H}(\theta_l)\right)[1,j]\right\},
		 \end{align}
	\begin{align}
	&\boldsymbol{J}_{\boldsymbol{\beta}_{l},\boldsymbol{\beta}_{l}} = \frac{2}{\sigma_{s}^{2}}\mathrm{Re}\left\{ \left([1,j] \otimes \mathrm{vec}\left(\boldsymbol{E}_{l}\boldsymbol{\Phi}_{l}\boldsymbol{G}_{l}\boldsymbol{X}\right)\right)^{H}[1,j] \right.\nonumber\\
	&\left.\otimes \mathrm{vec}\left(\boldsymbol{E}_{l}\boldsymbol{\Phi}_{l}\boldsymbol{G}_{l}\boldsymbol{X}\right)\right\} \nonumber\\
	&=\frac{2T}{\sigma_{s}^{2}}\mathrm{Re}\left\{ [1,j]^{H}[1,j]\left(  \mathrm{vec}\left(\boldsymbol{E}_{l}\boldsymbol{\Phi}_{l}\boldsymbol{G}_{l}\boldsymbol{X}\right)\right)^{H}\right.\nonumber\\  &\left.\cdot\mathrm{vec}\left(\boldsymbol{E}_{l}\boldsymbol{\Phi}_{l}\boldsymbol{G}_{l}\boldsymbol{X}\right)\right\} \nonumber\\ &=\frac{2T}{\sigma_{s}^{2}}\mathrm{tr}\left({\boldsymbol{E}}_{l}(\theta_l)\boldsymbol{\Phi}_{l}\boldsymbol{G}_{l}\boldsymbol{R}_{x}\boldsymbol{G}_{l}^{H}\boldsymbol{\Phi}_{l}^{H}{\boldsymbol{E}}_{l}^{H}(\theta_l)\right)\boldsymbol{I}_{2}.
	\end{align}
Therefore, Lemma \ref{FIM_lem} is finally proved.
	\section{Proof of Proposition \ref{CRB_lem}}\label{CRB_lem_proof}
	First, by substituting  \eqref{J_theta}-\eqref{J_beta} into \eqref{CRB_theta1}, we obtain \eqref{CRB_theta}. Then,
	we substitute \eqref{E_dot} into \eqref{J_theta}-\eqref{J_beta} to obtain the following results: 
	\begin{align}
	&\dot{\boldsymbol{E}}_{l}(\theta_l)\boldsymbol{\Phi}_{l}\boldsymbol{G}_{l}\boldsymbol{R}_{x}\boldsymbol{G}_{l}^{H}\boldsymbol{\Phi}_{l}^{H}\dot{\boldsymbol{E}}_{l}^{H}(\theta_l) = \nonumber\\
	& {\frac{4{\pi ^2}d_s^2{{\cos }^2}(\theta_l )}{ \lambda^2}} \left(\boldsymbol{D}_{N_s}\tilde{\boldsymbol{a}}_{l}(\theta_l)\boldsymbol{a}_{l}^{T}(\theta_l)+\tilde{\boldsymbol{a}}_{l}(\theta_l)\boldsymbol{a}_{l}^{T}(\theta_l)\boldsymbol{D}_{N}\right)\nonumber\\ 
	&\cdot\boldsymbol{\Phi}_{l}\boldsymbol{G}_{l}\boldsymbol{R}_{x}\boldsymbol{G}_{l}^{H}\boldsymbol{\Phi}_{l}^{H} \left(\boldsymbol{a}_{l}^{*}(\theta_l)\tilde{\boldsymbol{a}}_{l}^{H}(\theta_l)\boldsymbol{D}_{N_s}+\boldsymbol{D}_{N}\boldsymbol{a}_{l}^{*}(\theta_l)\tilde{\boldsymbol{a}}_{l}^{H}(\theta_l)\right)\nonumber\\
	&={\frac{4{\pi ^2}d_s^2{{\cos }^2}(\theta_l )}{\lambda^2}}\left(\frac{{\left( {{N_s} - 1} \right){N_s}\left( {2{N_s} - 1} \right)}}{6}{\mathop{\rm tr}\limits} \left( {{\boldsymbol{\phi}_{l}^{T}}{\boldsymbol{U}_l}\boldsymbol{\phi}_{l}^{*}} \right)\right. \nonumber\\
	&+ \frac{{\left( {{N_s} - 1} \right){N_s}}}{2}{\mathop{\rm tr}\limits} \left( {{\boldsymbol{\phi}_{l}^{T}}{\boldsymbol{U}_l}{{\boldsymbol{D}}_N}\boldsymbol{\phi}_{l}^{*}} \right)+ {N_s}{\rm{tr}}\left( {{\boldsymbol{\phi}_{l}^{T}}{{\boldsymbol{D}}_N}{\boldsymbol{U}_l}{{\boldsymbol{D}}_N}\boldsymbol{\phi}_{l}^{*}} \right) \nonumber \\
	&\left. + \frac{{\left( {{N_s} - 1} \right){N_s}}}{2}{\rm{tr}}\left( {{\boldsymbol{\phi}_{l}^{T}}{{\boldsymbol{D}}_N}{\boldsymbol{U}_l}\boldsymbol{\phi}_{l}^{*}} \right) \right), \label{appedixB_eq1}\\
	&{\boldsymbol{E}}_{l}(\theta_l)\boldsymbol{\Phi}_{l}\boldsymbol{G}_{l}\boldsymbol{R}_{x}\boldsymbol{G}_{l}^{H}\boldsymbol{\Phi}_{l}^{H}\dot{\boldsymbol{E}}_{l}^{H}(\theta_l)= \nonumber\\
	&j \frac{2\pi}{\lambda}d_s \cos (\theta_l)\tilde{\boldsymbol{a}}_{l}(\theta_l)\boldsymbol{a}_{l}^{T}(\theta_l) \boldsymbol{\Phi}_{l}\boldsymbol{G}_{l}\boldsymbol{R}_{x}\boldsymbol{G}_{l}^{H}\boldsymbol{\Phi}_{l}^{H}\nonumber\\
	&\cdot \left(\boldsymbol{a}_{l}^{*}(\theta_l)\tilde{\boldsymbol{a}}_{l}^{H}(\theta_l)\boldsymbol{D}_{N_s}+\boldsymbol{D}_{N}\boldsymbol{a}_{l}^{*}(\theta_l)\tilde{\boldsymbol{a}}_{l}^{H}(\theta_l)\right) \nonumber\\
	&=j \frac{2\pi}{\lambda}d_s \cos (\theta_l)\left(\frac{{\left( {{N_s} - 1} \right){N_s}}}{2}{\mathop{\rm tr}\limits} \left( {{\boldsymbol{\phi}_{l}^T}{\boldsymbol{U}_l} \boldsymbol{\phi}_{l}^{*}} \right) \right.\nonumber\\
	&\left. + {N_s}{\mathop{\rm tr}\limits} \left( {{\boldsymbol{\phi}_{l}^T}{\boldsymbol{U}_l}{{\boldsymbol{D}}_N}\boldsymbol{\phi}_{l}^{*}} \right)\right),\label{appedixB_eq2}\\
	&{\boldsymbol{E}}_{l}(\theta_l)\boldsymbol{\Phi}_{l}\boldsymbol{G}_{l}\boldsymbol{R}_{x}\boldsymbol{G}_{l}^{H}\boldsymbol{\Phi}_{l}^{H}{\boldsymbol{E}}_{l}^{H}(\theta_l) =\nonumber\\ &\tilde{\boldsymbol{a}}_{l}(\theta_l)\boldsymbol{a}_{l}^{T}(\theta_l)\boldsymbol{\Phi}_{l}\boldsymbol{G}_{l}\boldsymbol{R}_{x}\boldsymbol{G}_{l}^{H}\boldsymbol{\Phi}_{l}^{H}  \boldsymbol{a}_{l}^{*}(\theta_l)\tilde{\boldsymbol{a}}_{l}^{H}(\theta_l) \nonumber\\
	&= N_s{\mathrm{tr}}\left( {\boldsymbol{\phi}_{l}^T{\boldsymbol{U}_l}\boldsymbol{\phi}_{l}^*}\right),\label{appedixB_eq3}
\end{align}
where ${\boldsymbol{U}_l} = {{{\boldsymbol{A}}_l}{{\boldsymbol{G}}_{l}}{{\boldsymbol{R}}_x}{\boldsymbol{G}}_{l}^H{\boldsymbol{A}}_l^H}$ and  $\boldsymbol{A}_k = \mathrm{diag} (\boldsymbol{a}_l(\theta_l))$. Substituting \eqref{appedixB_eq1}-\eqref{appedixB_eq3} into \eqref{CRB_theta} yields Proposition \ref{CRB_lem}.
\section{Proof of Proposition \ref{CRB_limit_1}}\label{CRB_limit_1_proof}
Based on the closed-form CRB in \eqref{CRB_theta2} and by considering $N_s$ to be sufficiently large, we have
\begin{align}
	&\mathop {\lim }\limits_{{N_s} \to \infty } N_s^{3}\mathrm{CRB}_{l}(\theta_l) \nonumber\\
	&= \frac{N_s^{3} \sigma_{\mathrm{s}}^2 \lambda^2}{8 T|\beta_l|^2\pi^2 d_s^2 \cos^2(\theta_l)\left(\frac{{\left( {{N_s} - 1} \right){N_s}\left( {{N_s} + 1} \right)}}{{12}}{\mathop{\rm tr}\limits} \left( {\boldsymbol{\phi}_{l}^{T}{\boldsymbol{U}_l}\boldsymbol{\phi}_{l}^{*}} \right) \right)} \nonumber\\
	&= \frac{3\sigma_{\mathrm{s}}^2 \lambda^2}{2 T|\beta_l|^2\pi^2 d_s^2 \cos^2(\theta_l){\mathop{\rm tr}\limits} \left( {\boldsymbol{\phi}_{l}^{T}{\boldsymbol{U}_l}\boldsymbol{\phi}_{l}^{*}} \right) }.
\end{align}	
Therefore, $\mathop {\lim }\limits_{{N_s} \to \infty } N_s^{3}\mathrm{CRB}_{l}(\theta_l)$ is equal to a determined value independent of $N_s$. As a result, Proposition \ref{CRB_limit_1} is proved. 
\section{Proof of Lemma \ref{FIM_extended_lem}}\label{FIM_extended_lem_proof}
We first decompose the estimation parameter $\hat{\boldsymbol{h}}_{l}$ into real and imaginary parts that are denoted by $\mathrm{Re}\{\hat{\boldsymbol{h}}_{l}\}$ and $\mathrm{Im}\{\hat{\boldsymbol{h}}_{l}\}$, respectively. As a result, the FIM is equivalently expressed as
	\begin{align}
		\boldsymbol{F}_{l} = \left[\begin{array}{cc}
			\boldsymbol{F}_{\mathrm{Re}\{\hat{\boldsymbol{h}}_{l}\},\mathrm{Re}\{\hat{\boldsymbol{h}}_{l}\}}&\boldsymbol{F}_{\mathrm{Re}\{\hat{\boldsymbol{h}}_{l}\},\mathrm{Im}\{\hat{\boldsymbol{h}}_{l}\}}\\
			\boldsymbol{F}_{\mathrm{Im}\{\hat{\boldsymbol{h}}_{l}\},\mathrm{Re}\{\hat{\boldsymbol{h}}_{l}\}}& \boldsymbol{F}_{\mathrm{Im}\{\hat{\boldsymbol{h}}_{l}\},\mathrm{Im}\{\hat{\boldsymbol{h}}_{l}\}}
		\end{array}\right]. 
	\end{align}
	Then, we derive each part in the FIM as 
	\begin{align}
		&\boldsymbol{F}_{\mathrm{Re}\{\hat{\boldsymbol{h}}_{l}\},\mathrm{Re}\{\hat{\boldsymbol{h}}_{l}\}} \nonumber\\
		&= \frac{2}{\sigma_{{s}}^{2}}\mathrm{Re}\left\{\frac{\partial \hat{\boldsymbol{\eta}}_l^{H}}{\partial \mathrm{Re}\{\hat{\boldsymbol{h}}_{l}\}} \frac{\partial \hat{\boldsymbol{\eta}}_l}{\partial \mathrm{Re}\{\hat{\boldsymbol{h}}_{l}\}} \right\}\nonumber\\
		&=\frac{2}{\sigma_{{s}}^{2}}\mathrm{Re}\left\{\left((\boldsymbol{\Phi}_{l}\boldsymbol{G}_{l}\boldsymbol{X})^{T} \otimes \boldsymbol{I}_{N_s}\right)^{H} \left((\boldsymbol{\Phi}_{l}\boldsymbol{G}_{l}\boldsymbol{X})^{T} \otimes \boldsymbol{I}_{N_s}\right) \right\}\nonumber\\
		&=\frac{2}{\sigma_{{s}}^{2}}\mathrm{Re}\left\{(\boldsymbol{\Phi}_{l}\boldsymbol{G}_{l}\boldsymbol{X})^{*} (\boldsymbol{\Phi}_{l}\boldsymbol{G}_{l}\boldsymbol{X})^{T} \otimes \boldsymbol{I}_{N_s} \right\}, \\
		&\boldsymbol{F}_{\mathrm{Im}\{\hat{\boldsymbol{h}}_{l}\},\mathrm{Im}\{\hat{\boldsymbol{h}}_{l}\}} \nonumber\\
		&= \frac{2}{\sigma_{{s}}^{2}}\mathrm{Re}\left\{\frac{\partial \hat{\boldsymbol{\eta}}_l^{H}}{\partial \mathrm{Im}\{\hat{\boldsymbol{h}}_{l}\}} \frac{\partial \hat{\boldsymbol{\eta}}_l}{\partial \mathrm{Im}\{\hat{\boldsymbol{h}}_{l}\}} \right\}\nonumber\\
		&=\frac{2}{\sigma_{{s}}^{2}}\mathrm{Re}\left\{\left((\boldsymbol{\Phi}_{l}\boldsymbol{G}_{l}\boldsymbol{X})^{T} \otimes \boldsymbol{I}_{N_s}\right)^{H} \left((\boldsymbol{\Phi}_{l}\boldsymbol{G}_{l}\boldsymbol{X})^{T} \otimes \boldsymbol{I}_{N_s}\right) \right\}\nonumber\\
		&=\frac{2}{\sigma_{{s}}^{2}}\mathrm{Re}\left\{(\boldsymbol{\Phi}_{l}\boldsymbol{G}_{l}\boldsymbol{X})^{*} (\boldsymbol{\Phi}_{l}\boldsymbol{G}_{l}\boldsymbol{X})^{T} \otimes \boldsymbol{I}_{N_s} \right\}, \\
		&\boldsymbol{F}_{\mathrm{Re}\{\hat{\boldsymbol{h}}_{l}\},\mathrm{Im}\{\hat{\boldsymbol{h}}_{l}\}}\nonumber\\
		&=-\boldsymbol{F}_{\mathrm{Im}\{\hat{\boldsymbol{h}}_{l}\},\mathrm{Re}\{\hat{\boldsymbol{h}}_{l}\}} = \frac{2}{\sigma_{{s}}^{2}}\mathrm{Im}\left\{\frac{\partial \hat{\boldsymbol{\eta}}_l^{H}}{\partial \mathrm{Re}\{\hat{\boldsymbol{h}}_{l}\}} \frac{\partial \hat{\boldsymbol{\eta}}_l}{\partial \mathrm{Im}\{\hat{\boldsymbol{h}}_{l}\}} \right\}\nonumber\\
		&=\frac{2}{\sigma_{{s}}^{2}}\mathrm{Re}\left\{\left((\boldsymbol{\Phi}_{l}\boldsymbol{G}_{l}\boldsymbol{X})^{T} \otimes \boldsymbol{I}_{N_s}\right)^{H} \left((\boldsymbol{\Phi}_{l}\boldsymbol{G}_{l}\boldsymbol{X})^{T} \otimes \boldsymbol{I}_{N_s}\right) \right\}\nonumber\\
		&=\frac{2}{\sigma_{{s}}^{2}}\mathrm{Re}\left\{(\boldsymbol{\Phi}_{l}\boldsymbol{G}_{l}\boldsymbol{X})^{*} (\boldsymbol{\Phi}_{l}\boldsymbol{G}_{l}\boldsymbol{X})^{T} \otimes \boldsymbol{I}_{N_s} \right\}.
	\end{align}
As a result, the Lemma \ref{FIM_extended_lem} is obtained.
\section{Proof of Proposition \ref{CRB_limit_2}}\label{CRB_limit_2_proof}
Based on the closed-form CRB in \eqref{CRB_extend} and by considering $N_s$ to be sufficiently large, we have 
\begin{align}
	\mathop {\lim }\limits_{{N_s} \to \infty } \frac{\widetilde{\mathrm{CRB}}_l(\hat{\boldsymbol{E}}_{l})}{N_s} &= \frac{\sigma_{{s}}^2}{T}\mathrm{tr}\left((\boldsymbol{G}_{l}^{*}\boldsymbol{R}_{x}^{*}\boldsymbol{G}_{l}^{T})^{-1}\right).
\end{align}
Thus, $\mathop {\lim }\limits_{{N_s} \to \infty } \frac{\widetilde{\mathrm{CRB}}_l(\hat{\boldsymbol{E}}_{l})}{N_s}$ is equal to a constant value independent of $N_s$. As a result, Proposition \ref{CRB_limit_2} is proved.
\section{Proof of Proposition \ref{prop1}}\label{prop1_proof}
		First, we have
\begin{align}
	&\frac{1}{\Gamma_{l,k}}  {{\tilde{\boldsymbol{h}}_{l,k}^{H}\overline{\boldsymbol{W}}_{l,k}^{\star} \tilde{\boldsymbol{h}}_{l,k}}}=\frac{1}{\Gamma_{l,k}}  {{\tilde{\boldsymbol{h}}_{l,k}^{H}\overline{\boldsymbol{W}}_{l,k} \tilde{\boldsymbol{h}}_{l,k}}} \nonumber\\
	&\geq \sum\limits_{l \in \mathcal{L} }{\sum\limits_{k' \in \mathcal{K}_{l}\backslash k} {{\tilde{\boldsymbol{h}}_{l,k}^{H}\overline{\boldsymbol{W}}_{k'} \tilde{\boldsymbol{h}}_{l,k} }} +\tilde{\boldsymbol{h}}_{l,k}^{H} \overline{\boldsymbol{R}}_{0}\tilde{\boldsymbol{h}}_{l,k}+ \sigma^2} \nonumber\\
	&= \sum\limits_{l \in \mathcal{L} } \sum\limits_{k' \in \mathcal{K}_{l}\backslash k} {{\tilde{\boldsymbol{h}}_{l,k}^{H}\overline{\boldsymbol{W}}_{k'}^{\star} \tilde{\boldsymbol{h}}_{l,k} }} +\tilde{\boldsymbol{h}}_{l,k}^{H} \overline{\boldsymbol{R}}_{0}^{\star}\tilde{\boldsymbol{h}}_{l,k}+ \sigma^2, \forall k.
\end{align}
As a result, the SINR constraint in \eqref{P1_I_cons1} are still satisfied with \eqref{prop1.2}-\eqref{prop1.3}. Furthermore, for any $\boldsymbol{z}\in \mathbb{C}^{M\times 1}$, it holds that
\begin{align}
	&\boldsymbol{z}^{H}\left(\overline{\boldsymbol{W}}_{l,k}-\overline{\boldsymbol{W}}_{l,k}^{\star} \right) \boldsymbol{z} \nonumber\\
	&  =\boldsymbol{z}^{H}\overline{\boldsymbol{W}}_{l,k}\boldsymbol{z} -  \left|\boldsymbol{z}^{H}\overline{\boldsymbol{W}}_{l,k} \tilde{\boldsymbol{h}}_{l,k} \right|^{2}\left(\tilde{\boldsymbol{h}}_{l,k}^{H}\overline{\boldsymbol{W}}_{l,k} \tilde{\boldsymbol{h}}_{l,k} \right)^{-1}.
\end{align} 
According to the Cauchy-Schwarz inequality, we have 
\begin{align}
	\left|\boldsymbol{z}^{H}\overline{\boldsymbol{W}}_{l,k} \tilde{\boldsymbol{h}}_{l,k} \right|^{2}\leq \left(\boldsymbol{z}^{H}\overline{\boldsymbol{W}}_{l,k}\boldsymbol{z}\right)\left(\tilde{\boldsymbol{h}}_{l,k}^{H}\overline{\boldsymbol{W}}_{l,k} \tilde{\boldsymbol{h}}_{l,k} \right),
\end{align}
and thus $\boldsymbol{z}^{H}\left(\overline{\boldsymbol{W}}_{l,k}-\overline{\boldsymbol{W}}_{l,k}^{\star} \right) \boldsymbol{z} \geq 0$ also holds. As a result, we have $\overline{\boldsymbol{W}}_{l,k}-\overline{\boldsymbol{W}}_{l,k}^{\star} \succeq 0$, i.e., $\overline{\boldsymbol{W}}_{l,k}-\overline{\boldsymbol{W}}_{l,k}^{\star}$ is positive semidefinite. Therefore, it follows that $\overline{\boldsymbol{R}}_{0}^{\star}$ is also positive semi-definite. As a result,  $\{\overline{\boldsymbol{W}}_{l,k}^{\star}\}$ and $\overline{\boldsymbol{R}}_{0}^{\star}$ are optimal to problem (P1.1). This completes the proof.
 	\end{appendices}	
	
	\bibliographystyle{IEEEtran}
	\bibliography{IEEEabrv,myref}

\end{document}